\documentclass[10pt,letter,final]{article}

\usepackage[backref,
            natbib
            ,style = numeric-comp
            ,maxnames = 2
            ,backend = bibtex
            ,sorting=none
            ]{biblatex}
  
\usepackage[T1]{fontenc} 
\usepackage{palatino,eulervm}
\usepackage[english]{babel} 
\usepackage{csquotes}
\usepackage{amsmath,amssymb,amsfonts,amsthm} 
\usepackage[table,dvipsnames*,svgnames]{xcolor}
\usepackage{fixltx2e}
\usepackage{showkeys}
\usepackage{bm} 
\usepackage[top=2.5cm, bottom=2.5cm, outer=3cm, inner=3cm, heightrounded, marginparwidth=2cm, marginparsep=0.5cm]{geometry} 
\usepackage[active]{srcltx}
\usepackage{extarrows}

\usepackage{braket}
\usepackage{mathrsfs}
\usepackage{marginnote}
\usepackage[utf8]{inputenc}

\numberwithin{equation}{section}

\makeatletter

\let\@authors\@empty
\let\@email\@empty
\let\@affiliationone\@empty
\let\@affiliationtow\@empty
\let\@pdfsubject\@empty
\let\@keywords\@empty
\let\@preprint\@empty

\providecommand{\pdfsubject}[1]{\gdef\@pdfsubject{#1}}
\providecommand{\keywords}[1]{\gdef\@keywords{#1}}
\renewcommand{\author}[1]{\ifx\@authors\@empty\toks@\expandafter{#1}\else\toks@\expandafter{\@authors, #1}\fi\edef\@authors{\the\toks@}}
\providecommand{\email}[1]{\ifx\@email\@empty\toks@\expandafter{#1}\else\toks@\expandafter{\@email, #1}\fi\edef\@email{\the\toks@}}
\providecommand{\affiliationone}[1]{\gdef\@affiliationone{#1}}
\providecommand{\affiliationtwo}[1]{\gdef\@affiliationtwo{#1}}
\providecommand{\preprint}[1]{\gdef\@preprint{#1}}

\makeatother

\usepackage[hyperindex,breaklinks]{hyperref}

\hypersetup{
    unicode=false,          
    pdftoolbar=true,        
    pdfmenubar=true,        
    pdffitwindow=false,     
    pdfstartview={FitH},    
    pdfproducer={SGA, BUWS}, 
    pdftitle={Soft Black Hole Absorption Rates as Ward Identities},
    pdfsubject={Absorption rates of black holes and large gauge transformations},
    pdfkeywords={Black Holes, Absorption, LGT},
    pdfnewwindow=true,      
    colorlinks=true,       
    linkcolor=black,          
    citecolor=DarkGreen,        
    filecolor=magenta,      
    urlcolor=cyan!30!black!70         
}

\usepackage{graphicx}

\allowdisplaybreaks[3]
\newcommand{\bal}{\begin{align}}
\newcommand{\eal}{\end{align}}

\renewcommand{\[}{\begin{equation}}
\renewcommand{\]}{\end{equation}}

\bibliography{greybody.bib}

\begin{document}

\title{Greybody Factor in AdS Black Brane Spectroscopy: A Study with Scalar and Electromagnetic Perturbations}

\author{Atanu\ Bhatta\textsuperscript{a}}
\email{\href{mailto:atanu.bhatta@dituniversity.edu.in}{atanu.bhatta@dituniversity.edu.in}}

\affiliationone{%
\textsuperscript{a}Department of Physics, School of Physical Sciences, DIT University\\Mussoorie-Diversion Road, Makkawala, Dehradun, Uttarakhand -248009, India\\ 
\vspace{3mm}
}

\author{Arpit\ Maurya\textsuperscript{b}}
\email{\href{mailto:arpit.20phz0009@iitrpr.ac.in}{arpit.20phz0009@iitrpr.ac.in}}

\affiliationtwo{%
\textsuperscript{b}Department of Physics, Indian Institute of Technology Ropar \\ Rupnagar, Punjab, India 140001 \\
\vspace{3mm}
}

\keywords{Black holes, Absorption, LGT}
\pdfsubject{Black holes, Absorption, LGT}


\makeatletter
\thispagestyle{empty}

\begin{flushright}
\begingroup\ttfamily\@preprint\par\endgroup
\end{flushright}

\begin{centering}
\begingroup\Large\normalfont\bfseries\@title\par\endgroup
\vspace{1cm}

\begingroup\@authors\par\endgroup
\vspace{5mm}

\begingroup\itshape\@affiliationone\par\endgroup
\vspace{3mm}
\begingroup\itshape\@affiliationtwo\par\endgroup
\vspace{3mm}

\begingroup\ttfamily\@email\par\endgroup
\vspace{0.25cm}

\begin{minipage}{16cm}
\begin{abstract}
We compute the greybody factor for scalar and photon radiations in AdS Schwarzschild black brane. We consider the linearized field equations satisfied by the minimally coupled massive scalar field and the transverse components of the Maxwell fields on the black brane background and recast them to the Heun's equations by appropriate change of variables and field redefinitions. At the asymptotic region, we consider appropriate linear combinations of the normalizable and non-normalizable solutions of the field equations to construct in/outgoing modes. We consider the ingoing solutions near horizon to construct the IN state and compute the coefficients of the in- and outgoing modes of the asymptotic solutions in terms of the connection coefficients of the Heun functions. Finally by calculating ratio of the coefficient of outgoing modes to the ingoing modes we find an exact expression for the greybody factor. We find that the greybody factor for the transverse Maxwell fields and that for the scalars dual to the conformal scalar primaries with integer scaling dimensions vanish off. 
\end{abstract}
\end{minipage}

\vspace{2mm}
\rule{\textwidth}{.5mm}
\vspace{-1cm}

\end{centering}

\makeatother

\tableofcontents

\vspace{2mm}
\rule{\textwidth}{.5mm}
\vspace{-1cm}

%
%
\section{Introduction} \label{introduction}
The Hawking radiation~\cite{Hawking:1975vcx} from a black hole turns out to be a blackbody radiation when it is observed at the event horizon of the black hole. However, as it propagates outward and reaches to the observer at the asymptotic boundary the spectrum is modified by the non-trivial curved spacetime geometry, leading to a deviation from the pure blackbody spectrum. Specifically, at the asymptotic boundary the expectation value of the number of particles radiated with frequency $\omega$ becomes
\begin{eqnarray}
    \langle n (\omega) \rangle  = \frac{\gamma(\omega)}{e^{\frac{\omega}{T_H}} \pm 1}
\end{eqnarray}
where $T_H$ is the Hawking temperature, and the relative sign in the denominator takes $+ (-)$ if the particles are fermions (bosons). It is clear from the above expression that when $\gamma(\omega)$ is constant the spectrum of the Hawking radiation resembles the blackbody radiation. On the other hand, when $\gamma(\omega)$ is non-trivial, the radiation deviates from the blackbody radiation and is termed the greybody radiation. Thus $\gamma(\omega)$ quantifies the deviation and is referred to as the greybody factor. 

Now, some part of the emitted Hawking radiation get reflected by the surrounding geometry and reabsorbed by the black hole. Hence to compute the greybody factor of a given black hole one has to study the scattering of particles off the black holes. The scattering problem is realized via linearized field equations corresponding to the appropriate fields minimally coupled to the black hole background~\cite{Starobinsky:1973aij,Starobinskil:1974nkd,Page:1976df,Unruh:1976fm,Sanchez:1977si,Andersson:2000tf}. The field equation becomes Schr\"{o}dinger like equation when written in tortoise coordinate $r_*$ (which describe the spacetime geometry outside the black hole) defined as
\begin{eqnarray}
    r_* = \int \frac{dr}{f(r)}
\end{eqnarray}
where $f(r)$ is the blackening factor related to the $tt$-component of the black hole metric as $f(r) = -g_{tt}$. Specifically for a generic field $\Psi$, the field equation takes the following form:
\begin{eqnarray}
      \label{eq:schrodinger}
    - \frac{d^2 \Psi}{d r_*^2} + V(r_*) \Psi = \omega^2 \Psi
\end{eqnarray}
The explicit formulae for the potential $V(r_*)$ for scalar, vector and tensor fields on various black hole background are listed in~\cite{Natario:2004jd}. The greybody factor is obtained by computing the tunneling probability for the potential barrier as
\begin{eqnarray}
    \gamma(\omega) = |T(\omega)|^2
\end{eqnarray}
where $T(\omega)$ is the transmission coefficient for the potential considered. 

Although the computation of greybody factor is straightforward for the black holes in asymptotically flat and de-Sitter spacetime by the aforementioned prescription, it is more subtle for the case of anti de-Sitter (AdS) black holes. Due to the presence of reflecting asymptotic boundary, it is difficult to define ingoing and outgoing modes in the asymptotic region of the AdS space. Naively, the large AdS black holes absorb the emitted radiation after it is reflected by the asymptotic boundary and achieve a thermodynamically stable equilibrium with the gas of particles in its surrounding. Thus the large black holes in asymptotically AdS space do not evaporate. According to the AdS/CFT correspondence~\cite{Maldacena:1997re,Gubser:1998bc,Witten:1998qj}, the large AdS black holes provides a dual gravity description of a thermal state in the conformal field theory (CFT) living on the asymptotic boundary of the AdS black hole~\cite{Witten:1998zw}. However, it should be mentioned here that despite of the difficulty in defining the S-matrix in asymptotically AdS black holes, the greybody factor plays an important role in the AdS/CFT correspondence. For example, by computing greybody factor it is shown that the emission rate of AdS black holes precisely matches to that of their corresponding D-brane systems~\cite{Das:1996wn,Maldacena:1996ix}. The greybody factors for the asymptotically AdS black holes have been calculated directly by solving the field equations in some particular ranges of parameters with appropriate approximations of the potential function~\cite{Harmark:2007jy,Rocha:2009xy,Jorge:2014kra}, and as well as using the gauge/gravity correspondence~\cite{Teo:1998dw,Muller-Kirsten:1998ijv}\footnote{The greybody factor for small black holes in asymptotically AdS black hole has been computed in~\cite{BarraganAmado:2024tfu}. The small AdS black holes are thermodynamically unstable and evaporates. However there exists some ideas regarding dual description of such small black holes~\cite{Asplund:2008xd}}. An exact form of the greybody factor for AdS black holes has been out of reach until recently. In~\cite{Noda:2022zgk} the authors have provided an exact result in five-dimensional AdS-Kerr spacetime by using the Heun/gravity correspondence.

The linearized field equations on the black hole backgrounds can be recast into the Heun's equation, a linear second order differential equation with four regular singularities~\cite{ronveaux1995heun,fiziev2015heunfunctionsmodernpowerful}, by making appropriate change of variables and field redefinitions~\cite{Musiri:2003rs,BarraganAmado:2018zpa,Amado:2021erf,https://doi.org/10.1155/2018/8621573}. Two of the four singularities correspond to the positions of the outer horizon and the asymptotic boundary of the AdS black holes. The solutions around each singularity are related to the solutions around others via connection formulae~\cite{Bonelli:2022ten,Lisovyy_2022}. The analytic properties of the Heun functions have been used to study the scattering problems in AdS black hole backgrounds~\cite{Novaes:2014lha,Amado:2020zsr,BarraganAmado:2021uyw}. The Heun/gravity correspondence has been further used to compute the quasi-normal modes of black holes~\cite{Aminov:2020yma,Bianchi:2021xpr,Bonelli:2021uvf,Bianchi:2021mft}, and to investigate the correlators of scalar primary operators in thermal CFTs in a variety of settings, including in the thermal CFTs with and without chemical potentials~\cite{Dodelson:2022yvn,Bhatta:2022wga},  in the two-sided holography~\cite{Dodelson:2023vrw}, in the thermal CFTs on $R^1 \times H^3$~\cite{Bhatta:2023qcl}, the current-current correlators in thermal CFTs~\cite{He:2023wcs,Jia:2024zes}, etc.

It is worth mentioning here that the Heun's equation appears in the semi-classical limit of the BPZ equation satisfied by the five-point correlation function with one degenerate operator insertion in the Liouville CFT~\cite{Belavin:1984vu,Piatek:2017fyn,Bonelli:2022ten}. The crossing symmetry of the CFT correlators is used to find the relevant connection formulae among the solutions of the Heun's equation near different singular points~\cite{Bonelli:2022ten}. The Liouville CFT has a dual description via AGT correspondence as the BPS sector of $\mathcal{N} =2,\, SU(2)$ super-symmetric Yang-Mills theory in the $\Omega$-background~\cite{Alday:2009aq,Alday:2009fs}. Specifically, insertions of degenerate operators in the correlators of the Liouville CFT correspond to the insertions of the surface operators in the gauge theory and the semi-classical limit of the Liouville CFT corresponds to the Nekrasov-Shatashvili limit on the parameters of the $\Omega$-background~\cite{Maruyoshi:2010iu,Piatek:2017fyn}. In a word, the two-dimensional Liouville CFT serves as an auxiliary theory which is used in the study of the Heun/gravity correspondence.

In~\cite{Noda:2022zgk}, the authors used the connection formulae among the solutions of the Heun's equation near different singular points to obtain an exact expression of the greybody factor for five-dimensional Kerr-AdS black hole. In particular, the linearized equation for a massive scalar field minimally coupled to the black hole background has been written as the Heun's equation.  The scalar field equation admits two solutions near the black hole horizon, namely, the ingoing and outgoing solutions. One chooses the ingoing solution to be the boundary condition at the horizon and discards the other. Again, near the asymptotic boundary, the field equation admits two solutions: the normalizable mode, which falls off, and the non-normalizable mode, which diverges. The ingoing solution near the horizon is related to the normalizable and non-normalizable modes by the connection formulae. Then by considering appropriate linear combinations of the normalizable and non-normalizable modes one constructs the ingoing and outgoing modes near the asymptotic region as it is done using far region approximation method in~\cite{Jorge:2014kra}. Finally using the connection formulae among the Heun functions, an exact result for the greybody factor is obtained. 

In this paper we apply the techniques developed in~\cite{Noda:2022zgk} to analyze the spectroscopy of a black brane in asymptotically AdS space. Note that, the spectroscopy of the AdS black brane has been studied in~\cite{Rocha:2009xy} with massless minimally coupled scalar field. In this work, we consider a minimally coupled massive scalar field of mass $m$ propagating through the black brane in five-dimensional asymptotically AdS space. According to the AdS/CFT dictionary, the scalar field is dual to a scalar primary operator in the boundary CFT where the mass of the scalar field is related to the scaling dimension $\Delta$ of the primary operator as $m^2 = \Delta(\Delta -4)$\footnote{We have set the AdS radius $\ell = 1$.}. However, we recast the linearized equation of motion satisfied by the bulk scalar field into the Heun's equation by making appropriate change of variables and field redefinitions. Finally, we use the connection formulae among the solutions of the Heun's equation near the horizon and that near the asymptotic boundary respectively and obtain an exact result for the greybody factor for this case. Remarkably, we observe that for the class of scalar fields for which the scaling dimensions of corresponding dual conformal primary operators are {\it integers}, the greybody factor vanishes, i.e.,
\begin{eqnarray}
    \gamma(\omega) = 0  \qquad \text{for} \qquad  \Delta =  n + 2
\end{eqnarray}
where $n$ is any positive integer $(n>0)$. The result can be interpreted as follows. The scalar radiation of mass $m_n= \sqrt{n^2-4}$ cannot overcome the potential barrier $V(r_*)$ as elucidated earlier in~\eqref{eq:schrodinger}. The AdS boundary behaves as a perfect reflector for these scalar modes. Note that for small AdS black hole it was found that the greybody factor for scalar radiation vanishes for certain critical frequencies~\cite{Harmark:2007jy,Jorge:2014kra,Sakalli:2022xrb}. Our results hold for the large AdS black holes and any frequency of the massive scalar modes with mass $m_n$.

In this work we also consider the Maxwell fields minimally coupled to the asymptotically AdS black brane background. The gauge/gravity duality dictates that the transverse components of the Maxwell fields contribute to the source term of the global $U(1)$ conserved current operators of the boundary thermal CFT~\cite{Mueck:1998iz,Liu:1998ty,DHoker:1998bqu}. Again transforming the Maxwell's equations into the Heun's equation and computing the appropriate connection formulae we found that the greybody factor for the transverse gauge fields propagating through the AdS black brane vanishes too. 

The outline of the paper is as follows: In section~\ref{scalar}, we consider a massive Klein-Gordon field minimally coupled to AdS-Schwarzschild black brane. We prepare the corresponding IN state and using connection formulae among the solutions of the Heun's equation, we compute the greybody factor. In section \ref{maxwelltheory}, we study Maxwell theory on AdS-Schwarzschild black brane in radial gauge. We decompose the gauge field into transverse component and parallel component and find the equations of motion for both the components separately. We transform the equations of motion of the Maxwell fields into normal form of Heun's equation by a suitable change of variables and field redefinition. We study the behaviours of the transverse gauge files near the asymptotic boundary and prepare the IN mode by taking appropriate linear combinations. Finally, using the connection formulae, we get the greybody factor to vanish off. We conclude in section \ref{conclusion}.
\section{Scalar fields in AdS-Schwarzschild black brane} \label{scalar}
According to AdS/CFT dictionary, scalar fields in the bulk contribute to the source term of the boundary scalar primary operator. We consider a minimally coupled scalar field $\Phi$ of mass $m$ in the five dimensional AdS Schwarzschild black brane background
\begin{equation}
    S_{\Phi} = \frac{1}{2} \int drd^4x \, \sqrt{-g}  \left[\partial_I \Phi \partial^I \Phi + m^2 \Phi \right]
\end{equation}
where $g$ is the determinant of the metric 
\begin{equation}
    \label{eq:metric}
    ds^2 = \frac{R^2}{r^2} \left( \frac{dr^2}{f(r)} - f(r)dt^2 + dx_i dx_i\right) \quad \text{with} \quad f(r) = 1- \frac{r^4}{r_h^4}
\end{equation}
Here we are using the following notation: $X^I \equiv (r, x^\mu) \equiv (r, t, {\bf x} ) \equiv (r, t, x_1, x_2, x_3)$. The event horizon is at $r=r_h$ and the asymptotic boundary is at $r\to 0$.

The boundary value of the scalar field acts as a source term of the scalar primary operator $\mathcal{O}$ in the boundary CFT
\begin{equation}
    \label{eq:cft_action}
    S_{CFT} = S_0 + \int d^4x ~\sqrt{-\gamma} \phi \mathcal{O} ; 
\end{equation}
where $\gamma$ is the determinant of the induced metric on the conformal boundary of the AdS space and 
\begin{equation}
   \lim_{r \to 0} \Phi (r, x) = \phi (x) \quad \text{(up to a normalization factor)}
\end{equation}
The mass of the bulk scalar field is related to the scaling dimension of the boundary scalar primary operator 
\begin{equation}
    m^2 = \Delta(\Delta -4)  
\end{equation}

The equation of motion for the bulk scalar field 
\begin{equation}
    \frac{1}{\sqrt{-g}} \partial_I \left( \sqrt{-g} g^{\IJ} \partial_J \Phi \right) -m^2 \Phi = 0
\end{equation}
reads
\begin{equation} \label{eom}
    f(r) \partial_r^2 \Phi - \frac{1}{r} \left( 3f(r)- r f'(r) \right) \partial_r \Phi + \left( - \frac{1}{f(r)}\partial_t^2 + \partial^2 -\frac{m^2}{r^2}\right) \Phi = 0
\end{equation}
Considering the time translation and space translation symmetries in the metric, we choose the  ansatz for the massive scalar field as following
\begin{equation}
    \Phi(t,r,{\bf{x}})=\int d\omega \int d^3{\bf{x}} \ e^{-i\omega t} e^{-i {\bf{k}} \cdot {\bf{x}}} \phi_{\omega,{\bf{k}}}(r) 
\end{equation}
Substituting this ansatz into \eqref{eom}, we obtain the radial equation given by
\begin{equation} \label{eomr}
    \bigg( f(r) \partial_r^2  - \frac{1}{r} \big( 3f(r)- r f'(r) \big) \partial_r  + \left(  \frac{\omega^2}{f(r)} - {\bf{k}}^2 -\frac{m^2}{r^2}\right) \bigg)\phi_{\omega,{\bf{k}}}(r)  = 0
\end{equation}
In order to transform the above radial equation into normal form of Heun's equation, it is convenient to redefine the radial coordinate and field via transformations given by~\cite{Bhatta:2023qcl}
\begin{equation} \label{scalar_heunz}
z =\frac{r_h^2}{r^2+r_{h}^2}
\end{equation}
\begin{equation} \label{scalar_field_trans}
\phi_{\omega,{\bf{k}}}(r) =\left( \frac{f(r)}{r^3} \frac{d z}{d r}\right)^{-1 / 2} \psi_{\omega,{\bf{k}}}(z)
\end{equation}
Consequently, we end up with the Heun's equation
\begin{equation}\label{heun}
\left(\partial_z^2+\frac{\frac{1}{4}-a_1^2}{(z-1)^2}-\frac{\frac{1}{2}-a_0^2-a_1^2-a_h^2+a_\infty^2+u}{z(z-1)}+ \frac{\frac{1}{4}-a_h^2}{(z-z_h)^2} + \frac{u}{z(z-z_h)}+ \frac{\frac{1}{4}-a_0^2}{z^2} \right) \psi_{\omega,{\bf{k}}}(z) = 0
\end{equation} 
with the Heun's parameters given by
\begin{equation}\label{scalar_heunpar}
   z_h=\frac{1}{2}, \quad \quad a_0=0, \quad \quad  a_h=\frac{i r_h \omega}{4},  \quad \quad  a_1=\frac{\Delta-2}{2}, \quad \quad a_{\infty}=\frac{ r_h \omega}{4}
\end{equation} 
and $u$ is defined as
\begin{equation}
    u=\frac{1}{8} \left(r_h^2 (\omega^2-2 {\bf{k}}^2)-2(\Delta-2)^2\right)
\end{equation}

\subsection{Boundary conditions at horizon} \label{bdry cond}
The Heun's equation~\eqref{heun} obtained in the last section, have four regular singular points located at $z=0,1,z_h$ and $\infty$, admits two linearly independent solutions around each regular singular point. We are interested in the in-going solutions at the horizon ($r=r_h$) of the AdS black brane studied in the tortoise coordinates. Since the horizon is mapped to $z=z_h=1/2$, we have to identify the corresponding solution of the Heun's equation around $z=z_h$.
In the tortoise coordinates 
\begin{equation}
    r_*=\int\frac{1}{f(r)} dr
\end{equation}
near the horizon $r=r_h$ the equations for $\phi_{\omega,{\bf{k}}}(r)$ take the following form
\begin{equation} 
   \bigg(  \partial^2_{r_*}    +  \omega^2  \bigg) \phi_{\omega,{\bf{k}}}(r_*) = 0
\end{equation}
There are two kinds of solutions to the above equation
\begin{eqnarray}
    \phi_{\omega,{\bf{k}}}(r_*) &\sim& e^{- i \omega r_*} \qquad \text{(in-going)} \\
                        &\sim& e^{ i \omega r_*} \qquad ~~\, \text{(out-going)}
\end{eqnarray}
 We recast the above solutions in the $r$-coordinate and write the general solution as a superposition of those 
\begin{equation}
  \left.  \phi_{\omega,{\bf{k}}}(r) \right|_{r=r_h}= C_1 \times (r-r_h)^{-\frac{i \omega}{f'(r_h)}} (1+ \cdots)+ C_2 \times (r-r_h)^{\frac{i \omega}{f'(r_h)}} (1 + \cdots )
\end{equation}
where $C_1$ and $C_2$ are integration constants and $f'(r) = df/dr$. We now impose the ingoing boundary condition at horizon demanding that anything can fall into the horizon but nothing can come out of it. This is achieved by setting $C_2=0$ and then we have
\begin{equation} \label{aperphorizon}
  \left.  \phi_{\omega,{\bf{k}}}(r) \right|_{r=r_h}= C_1 \times (r-r_h)^{-\frac{i \omega}{f'(r_h)}} (1 + \cdots) = C_1 \times (r-r_h)^{\frac{i \omega r_h}{4}} (1 + \cdots )
\end{equation} 
This is the desired behaviour of $\phi_{\omega,{\bf{k}}}(r)$ near horizon. 

Now we solve the Heun's equation \eqref{heun} around $z=z_h$ 
\begin{equation} \label{aizhor}
 \left. \psi_{\omega,{\bf{k}}}(z)  \right|_{z=z_h} = A (z_h-z)^{\frac{1}{2}+a_h}(1 +\dots ) + B (z_h-z)^{\frac{1}{2}-a_h}(1 +\dots )
\end{equation}
Substituting the expression~\eqref{scalar_heunpar} for $a_h$ and $z_h=1/2$ obtained in the previous section in the above solution and considering the change of variable~\eqref{scalar_heunz} and field redefinition~\eqref{scalar_field_trans} we recognize the in-going solution in the $z$-coordinate to be
\begin{equation} 
 \left. \psi_{\omega,{\bf{k}}}(z)  \right|_{z=z_h} \sim (z_h-z)^{\frac{1}{2}-a_h} 
\end{equation}
\subsection{Asymptotic behaviour} \label{asymptotic0}
Near the asymptotic boundary $r\to 0$ of the AdS black brane we observe that the scalar field satisfies the following equation
\begin{eqnarray}
\label{eq:scalar_ODE_near_bdy}
\left[r^2 \partial_r^2 - 3 r \partial_r + (\omega^2 - \mathbf{k}^2) r^2 - m^2 \right] \phi_{\omega,{\bf{k}}}(r)   = 0
\end{eqnarray}

We write the equation in $u = \omega r$ variable as
\begin{eqnarray}
\left[u^2 \partial_u^2 - 3 u\partial_u + \left(1 - \frac{\mathbf{k}^2}{\omega^2} \right) u^2 - m^2 \right] \phi_{\omega,{\bf{k}}}(u)   = 0
\end{eqnarray}
We identify the above equation as Bessel's equation. The above equation exactly matches with the equation (74) of~\cite{Noda:2022zgk} with the following identifications: $L=1$, $\mu = m$ and $\lambda = \mathbf{k}^2$. 

The linearly independent solutions can be written in terms of the Bessel (J) and Neumann (N) functions. The asymptotic behaviours near $u=0$ are as follows 
\begin{eqnarray}
u^2 J_{\Delta-2} \left( \sqrt{ 1 - \frac{\mathbf{k}^2}{\omega^2} } u\right) &\simeq&  A_J (\omega, \mathbf{k}) r^\Delta \\
u^2 N_{\Delta-2} \left( \sqrt{ 1 - \frac{\mathbf{k}^2}{\omega^2} } u\right) &\simeq&  A_N (\omega, \mathbf{k})  r^\Delta + B_N (\omega, \mathbf{k})  r^{4-\Delta}
\end{eqnarray}
where
\begin{eqnarray}
\label{eq:AJ}
A_J (\omega, \mathbf{k})   &=& \frac{4\left( 1 - \frac{\mathbf{k}^2}{\omega^2}\right)^{\frac{\Delta-2}{2}}}{\Gamma[\Delta-1]}\left( \frac{\omega}{2}\right)^\Delta, \\
\label{eq:AN}
A_N (\omega, \mathbf{k}) &=&  -\frac{\cos (\pi \Delta) \, \Gamma[2-\Delta]\, 2^{2-\Delta}}{\pi} \left( 1 - \frac{\mathbf{k}^2}{\omega^2}\right)^{\frac{\Delta-2}{2}} \omega^\Delta \\
\label{eq:BN}
B_N  (\omega, \mathbf{k}) &=& - \frac{\Gamma[\Delta-2]\,2^{\Delta-2}}{\pi} \left( 1 - \frac{\mathbf{k}^2}{\omega^2}\right)^{\frac{2-\Delta}{2}} \omega^{4-\Delta}
\end{eqnarray}
The solution, that behaves as $\simeq r^\Delta (\simeq r^{4-\Delta})$ near the asymptotic boundary, is identified as normalizable (non-normalizable) mode.

On the other hand, near $z=1$ which corresponds to the boundary limit in $z$-coordinate, the linearly independent solutions of the Heun's equation behave as follows
\begin{eqnarray}
     \psi_{\omega,{\bf{k}}}^{(11)}(z) \simeq (1-z)^{\frac{1}{2} + a_1} &\simeq& r_h^{1-\Delta} \, r^{\Delta -1} \\
     \psi_{\omega,{\bf{k}}}^{(12)}(z) \simeq (1-z)^{\frac{1}{2} - a_1} &\simeq& r_h^{\Delta - 3} \, r^{3-\Delta}
\end{eqnarray}
since $a_1 = (\Delta-2)/2$. Using~\eqref{scalar_heunz} and~\eqref{scalar_field_trans}, near the asymptotic boundary we find
\begin{eqnarray} \label{phi_decomposition}
    \phi_{\omega,{\bf{k}}}(r) \simeq \left( A_{11}  \, r^\Delta + \cdots \right) +  \left( A_{12} \, r^{4-\Delta} + \cdots \right)
\end{eqnarray}
where 
\begin{eqnarray}
    A_{11} = - \frac{i }{\sqrt{2}}r_h^{2-\Delta}, \qquad \text{and} \qquad A_{12} = - \frac{i}{\sqrt{2}} r_h^{\Delta-2}
\end{eqnarray}
\subsection{Connection formulae} 
Now, the linearly independent solutions of the Heun's equation~\eqref{heun} around each singularity are related to that around the other via connection coefficients. Specifically we relate the in-going solution $\psi_{\omega,{\bf{k}}}^{(z_h,\text{in})}(z)$ near $z=z_h$ to the solutions near $z=1$ as~\cite{Bonelli:2022ten,Dodelson:2022yvn,Consoli:2022eey}
\begin{eqnarray} 
    \psi_{\omega,{\bf{k}}}^{(z_h,\text{in})}(z) = C_{21} \psi_{\omega,{\bf{k}}}^{(11)}(z) + C_{22} \psi_{\omega,{\bf{k}}}^{(12)}(z)
\end{eqnarray}
where
\begin{eqnarray} \label{scalar_c21}
    C_{21} &=& z_h^{\frac{1}{2}-a_0-a_h} (1-z_h)^{a_h - a_1} e^{-\frac{1}{2}(\partial_{a_h} + \partial_{a_1})F} \sum_{\theta' = \pm} \mathcal{M}_{-\theta'} (a_h, a; a_0) \mathcal{M}_{(-\theta')+} (a,a_1; a_\infty) z_h^{\theta' a} e^{-\frac{\theta'}{2}\partial_a F} \cr 
    && 
\end{eqnarray}
and
\begin{eqnarray} \label{scalar_c22}
    C_{22} &=& z_h^{\frac{1}{2}-a_0-a_h} (1-z_h)^{a_h - a_1} e^{-\frac{1}{2}(\partial_{a_h} - \partial_{a_1})F} \sum_{\theta' = \pm} \mathcal{M}_{-\theta'} (a_h, a; a_0) \mathcal{M}_{(-\theta')-} (a,a_1; a_\infty) z_h^{\theta' a} e^{-\frac{\theta'}{2}\partial_a F} \cr 
    && 
\end{eqnarray}
with
\begin{eqnarray} 
    \mathcal{M}_{\theta\theta'} (a_0, a_1; a_2) = \frac{\Gamma(-2 \theta' a_1)}{\Gamma\left( \frac{1}{2} +\theta a_0 - \theta' a_1 + a_2\right)} \frac{\Gamma(1+ 2 \theta a_0)}{\Gamma\left( \frac{1}{2} +\theta a_0 - \theta' a_1 - a_2\right)}
\end{eqnarray}
In~\eqref{scalar_c21} and~\eqref{scalar_c22}, we have introduced a parameter $a$ which is related to the Heun's parameters via Matone relation~\cite{Matone:1995rx,Flume:2004rp}
\begin{eqnarray}
    u = -a^2 +a_h^2 -\frac{1}{4} + a_0^2 + z_h \frac{\partial F}{\partial z_h}
\end{eqnarray}
The function $F$\footnote{By leveraging the AGT correspondence the classical conformal block  can be identified
 with the Nekrasov Shatashvili partition 
function in four dimensional $\mathcal{N}=2$ $SU(2)$ gauge theory. This partition function is typically represented as a power series in \(z_h\), and the 
series is convergent only when \(|z_h| < 1\) \cite{Arnaudo:2022ivo}.} appearing in the above equation as well as in~\eqref{scalar_c21} and~\eqref{scalar_c22} is the classical conformal block which is related to the four point conformal blocks of the Liouville CFT in the semi-classical limit
\begin{eqnarray}
    \lim_{c \to \infty} \mathcal{F} (\Delta_0, \Delta_1, \Delta_h, \Delta_\infty; \Delta; z) \simeq z_h^{\Delta-\Delta_h-\Delta_0} ~ e^{\frac{c}{6} F(z)}
\end{eqnarray}
where $c$ is the Virasoro central charge of the Liouville CFT and the four primary operators with conformal dimensions $\Delta_0, \Delta_1, \Delta_h$, and $\Delta_\infty$ are inserted at $z = 0,\, 1,\, z_h$ and $\infty$ respectively (see Appendix~\ref{classical_block}).
\subsection{Greybody factor}
Let us consider the case when $m^2 \neq 0$. Then we can use the Heun functions to find the greybody factor of the black brane as discussed in~\cite{Noda:2022zgk}. 

Note that near the asymptotic boundary, the differential equation~\eqref{eq:scalar_ODE_near_bdy} is not that of Schr\"{o}dinger's type. So, the solutions do not look like $\exp (\pm i r)$ (even in tortoise coordinates). However we can define the net flux near the boundary as
\begin{eqnarray}
    \mathcal{F}_b  \propto \frac{i}{2}\left( \phi(r) \Pi^*(r) - \phi^*(r) \Pi (r)\right) = - \text{ Im}(\phi^* \Pi) 
\end{eqnarray}
where $\Pi$ is the canonical momentum field 
\begin{eqnarray}
    \Pi(r) = \sqrt{-g} g^{rr} \partial_r \phi (r)
\end{eqnarray}
Clearly, real solutions ($J_\nu$ and $N_\nu$ ) result in zero flux. Changing to complex combinations (Hankel basis) yields non-zero fluxes.
\begin{eqnarray}
    H^{(1)}_\nu  &=& J_\nu  + i N_\nu \cr 
    H^{(2)}_\nu  &=& J_\nu  - i N_\nu 
\end{eqnarray}
In particular, the two combinations, $H^{(1)}_\nu$ and $H^{(1)}_\nu$ have equal and opposite flux near the asymptotic boundary, i.e.,
\begin{eqnarray}
    \mathcal{F}_b \left(H^{(1)}_\nu \right) = - \mathcal{F}_b \left(H^{(2)}_\nu \right)
\end{eqnarray}
This relegates one of the such complex combinations to be called "in-going" and the other to be "out-going". 

Now, first we construct IN mode in the Heun basis as 
\begin{eqnarray} \label{scalar_in_mode_1}
   \psi_{\omega,{\bf{k}}}^{\text{IN}}(z) &=& \left\{ \begin{matrix} \psi_{\omega,{\bf{k}}}^{(z_h,\text{in})}(z) \simeq (z_h-z)^{\frac{1}{2} -a_h} \qquad \qquad \qquad ~z \to z_h\\ \\ C_{21} \psi_{\omega,{\bf{k}}}^{(11)}(z) + C_{22} \psi_{\omega,{\bf{k}}}^{(12)}(z)  \qquad \qquad \qquad \quad z \to 1 \end{matrix} \right. 
\end{eqnarray}
As elucidated earlier, $\psi_{\omega,{\bf{k}}}^{(11)}$ and $\psi_{\omega,{\bf{k}}}^{(12)}$ behaves as $\sim r^{\Delta-1}$ and $\sim r^{3-\Delta}$ respectively near the asymptotic boundary. Since we need in/out-going modes at the asymptotic boundary to compute the greybody factor, we change the basis of the IN mode by defining (see Appendix \ref{IN mode}) 
\begin{eqnarray}
    \psi_{\omega,{\bf{k}}}^+(z) &=& \frac{A_J + i A_N}{A_{11}} \psi_{\omega,{\bf{k}}}^{(11)}(z) + i \frac{B_N}{A_{12}} \psi_{\omega,{\bf{k}}}^{(12)}(z) \\
    \psi_{\omega,{\bf{k}}}^-(z) &=& \frac{A_J - i A_N}{A_{11}} \psi_{\omega,{\bf{k}}}^{(11)}(z) - i \frac{B_N}{A_{12}} \psi_{\omega,{\bf{k}}}^{(12)}(z)
\end{eqnarray}
such that the IN modes become
\begin{eqnarray} \label{scalar_in_mode_2}
   \psi_{\omega,{\bf{k}}}^{\text{IN}}(z)  &=& \left\{ \begin{matrix} \psi_{\omega,{\bf{k}}}^{(z_h,\text{in})}(z) \simeq (z_h-z)^{\frac{1}{2} -a_h} \qquad \qquad \qquad ~z \to z_h\\ \\ C_{\text{out}} \psi_{\omega,{\bf{k}}}^+(z) + C_{\text{in}} \psi_{\omega,{\bf{k}}}^-(z)  \qquad \qquad \qquad \quad z \to 1 \end{matrix} \right. 
\end{eqnarray}
The asymptotic forms of $\psi_{\omega,{\bf{k}}}^+$ and $\psi_{\omega,{\bf{k}}}^-$ are the outgoing and ingoing modes respectively. 

From~\eqref{scalar_in_mode_1} and~\eqref{scalar_in_mode_2} we obtain
\begin{eqnarray}
    C_{\text{out}} &=& \frac{A_{11}A_{12}}{2i A_J B_N} \left[ i \frac{B_N}{A_{12}} C_{21}- \frac{A_J+ i A_N}{A_{11}} C_{22} \right] \\
    C_{\text{in}} &=& \frac{A_{11}A_{12}}{2i A_J B_N} \left[ i \frac{B_N}{A_{12}} C_{21} + \frac{A_J - i A_N}{A_{11}} C_{22} \right]
\end{eqnarray}
Now, by construction 
\begin{eqnarray}
\mathcal{F}_b \left( \psi_{\omega,{\bf{k}}}^+ \right)  =  + \mathcal{F}_0; \qquad \qquad \mathcal{F}_b \left( \psi_{\omega,{\bf{k}}}^- \right)  =  - \mathcal{F}_0
\end{eqnarray}
with same positive normalization $\mathcal{F}_0$. Using linearity of the flux we find the net outward flux near the boundary to be
\begin{eqnarray}
 \mathcal{F}_{bdy} = \left| C_{out} \right|^2 \mathcal{F}_0 - \left| C_{in} \right|^2 \mathcal{F}_0
\end{eqnarray}
Since there is no source or sink between the horizon and the boundary and we have only ingoing modes at the horizon, the conservation of flux reads
\begin{eqnarray}
\mathcal{F}_{hor} + \mathcal{F}_{bdy} = 0
\end{eqnarray}
or,
\begin{eqnarray}
 \mathcal{F}_{hor}  = - \mathcal{F}_{bdy}
\end{eqnarray}
The greybody factor is defined as the fraction of ingoing boundary flux that is absorbed by the horizon
\begin{eqnarray}
  \gamma_{\mathbf{k}} (\omega) = \frac{\mathcal{F}_{hor}}{\mathcal{F}_{bdy}^{(in)}}
\end{eqnarray}
Since $ \mathcal{F}_{bdy}^{(in)} = |C_{in}|^2 \mathcal{F}_0 $, finally we obtain the greybody factor in terms of the connection coefficients of the Heun functions as
\begin{eqnarray}
    \gamma_{\mathbf{k}} (\omega) &=& 1 - \left| \frac{C_{\text{out}}}{C_{\text{in}}} \right|^2 \cr 
                                 &=& 1 -  \frac{\left| i \frac{B_N}{A_{12}} C_{21} - \frac{A_J + i A_N}{A_{11}} C_{22} \right|^2}{\left| i \frac{B_N}{A_{12}} C_{21} + \frac{A_J- i A_N}{A_{11}} C_{22} \right|^2}
\end{eqnarray}
\subsection{Scaling dimensions = positive integers}
Let us consider the case when $\Delta$ is positive integer, i.e.,
\begin{eqnarray}
    \Delta = n + 2 ; \qquad n \in \mathbb{Z^+}   
\end{eqnarray}
To systematically analyze this case, we regularize the order of the Bessel functions as
\begin{eqnarray}
    \nu = \Delta -2 = n + \epsilon, \qquad \epsilon \ll 1
\end{eqnarray}
Note that 
\begin{eqnarray}
    N_\nu (y) )= \cot (\pi \nu) J_\nu (y)- \frac{1}{\sin (\pi \nu)} J_{-\nu} (y) 
\end{eqnarray}
where $y = \left(\sqrt{\omega^2 - \bf{k}^2} \right)r$. Now 
\begin{eqnarray}
    \sin ( \pi \nu ) &=&  \sin (\pi n + \pi \epsilon) \approx (-1)^n \pi \epsilon \left( 1 - \frac{\pi^2 \epsilon^2}{6} \right)  \cr 
    \cos ( \pi \nu ) &=&  \cos (\pi n + \pi \epsilon) \approx (-1)^n  \left( 1 - \frac{\pi^2 \epsilon^2}{2} \right)
\end{eqnarray}
Hence,
\begin{eqnarray}
    \cot (\pi \nu) \approx \frac{1}{\pi \epsilon} - \frac{\pi \epsilon}{3}
\end{eqnarray}
and
\begin{eqnarray}
    \frac{1}{\sin (\pi \nu)} \approx \frac{(-1)^n}{\pi \epsilon}
\end{eqnarray}
Therefore, 
\begin{eqnarray}
    N_\nu (y) \approx \left( \frac{1}{\pi \epsilon} - \frac{\pi \epsilon}{3} \right) J_\nu (y) - \frac{(-1)^n}{\pi \epsilon} J_{-\nu} (y) 
\end{eqnarray}
So, as $\epsilon \to 0$, both $J_\nu$ and $J_{-\nu}$ both appear with $1/\epsilon$-prefactors.
Now,
\begin{eqnarray}
    J_{n+\epsilon} (y) &\approx& J_n(y) + \epsilon \left. \frac{\partial J_\nu}{\partial \nu} \right|_{\nu = n} \\
    J_{-n -\epsilon} (y) &\approx& J_{-n}(y) - \epsilon \left. \frac{\partial J_\nu}{\partial \nu} \right|_{\nu = -n}
\end{eqnarray}
Hence
\begin{eqnarray}
    N_{n+\epsilon} (y) \approx \frac{1}{\pi \epsilon} \left[ J_n (y) - (-1)^n J_{-n} (y)\right] + \frac{1}{\pi} \left[ \left. \frac{\partial J_\nu}{\partial \nu} \right|_{\nu = n} - (-1)^n \left. \frac{\partial J_\nu}{\partial \nu} \right|_{\nu = -n} \right] - \frac{\pi \epsilon}{3} J_n (y)
\end{eqnarray}
For integer order using the following identity
\begin{eqnarray}
    J_{-n} (y) = (-1)^n J_n (y) 
\end{eqnarray}
we obtain
\begin{eqnarray}
    N_{n+\epsilon} (y) &\approx& \frac{1}{\pi \epsilon} \left[ J_n (y) - (-1)^{2n} J_{n} (y)\right] + \frac{1}{\pi} \left[ \left. \frac{\partial J_\nu}{\partial \nu} \right|_{\nu = n} - (-1)^n \left. \frac{\partial J_\nu}{\partial \nu} \right|_{\nu =-n} \right] - \frac{\pi \epsilon}{3} J_n (y) \cr 
    &=& \frac{1}{\pi \epsilon} \left[ J_n (y) -  J_{n} (y)\right] + \frac{1}{\pi} \left[ \left. \frac{\partial J_\nu}{\partial \nu} \right|_{\nu = n} - (-1)^n \left. \frac{\partial J_\nu}{\partial \nu} \right|_{\nu = -n} \right] - \frac{\pi \epsilon}{3} J_n (y) \cr
    &\approx& \frac{1}{\pi} \left[ \left. \frac{\partial J_\nu}{\partial \nu} \right|_{\nu = n} - (-1)^n \left. \frac{\partial J_\nu}{\partial \nu} \right|_{\nu =- n} \right]
\end{eqnarray}
We find that the poles cancel at leading order.

Now as $y \to 0$ we have~\cite{abramowitz+stegun}
\begin{eqnarray}
    N_n (y) &=& \frac{2}{\pi} \ln (y/2) J_n (y) - \frac{(y/2)^{-n}}{\pi} \sum_{k=0}^{n-1} \frac{(n-k-1)!}{k!} \left( \frac{y^2}{4}\right)^k \cr 
   && \qquad 
   \qquad \qquad - \frac{(y/2)^n}{\pi} \sum_{k=0}^\infty [\psi_0 (k+1) + \psi_0(n + k +1) ] \frac{(-y^2/4)^k}{k! (n+k)!}
\end{eqnarray}
where $\psi_0$ is the digamma function, and
\begin{eqnarray}
    J_n(y) = \left( \frac{y}{2}\right)^2 \sum_{k=0}^\infty \frac{(-y^2/4)^k}{k! (n+k)!}
\end{eqnarray}
So, near the asymptotic boundary, to the leading order
\begin{eqnarray}
    N_n (y) &\sim& y^n [A + B \ln y ] + C y^{-n} \\
    J_n (y) &\sim& D y^n 
\end{eqnarray}
Thus we see that the $\epsilon$-regularization works appropriately with the Bessel functions and generate the desired logarithmic term for the integer scaling dimension.

Now for this case, let us calculate the net flux at the boundary as prescribed in the last section. In the Hankel basis, the ingoing boundary solution $(H^{(1)}_\nu)$ will take the following form
\begin{eqnarray}
    \phi_{\omega, \bf{k}} (r) \sim \mathcal{A} (\omega, {\bf k}) r^{\Delta} + \mathcal{B} (\omega, {\bf k}) r^{4-\Delta} + \mathcal{C} (\omega, {\bf k}) r^{\Delta} \ln r 
\end{eqnarray}
where
\begin{eqnarray}
    \mathcal{A} = D + i A; \quad \mathcal{B} = i C; \quad \mathcal{C} = iB
\end{eqnarray}
Complex conjugate of the field becomes
\begin{eqnarray}
    \phi^*_{\omega, \bf{k}} (r) \sim \mathcal{A}^* (\omega, {\bf k}) r^{n+2} + \mathcal{B}^* (\omega, {\bf k}) r^{2-n} + \mathcal{C}^* (\omega, {\bf k}) r^{n+2} \ln r 
\end{eqnarray}
and the derivative of the field
\begin{eqnarray}
    \partial_r \phi_{\omega, \bf k}(r) = \Delta \mathcal{A} r^{\Delta -1} + (4 - \Delta) \mathcal{B} r^{3- \Delta} + \Delta \mathcal{C} r^{\Delta - 1} \ln r + \mathcal{C} r^{\Delta -1}
\end{eqnarray}
Using these we compute the momentum field to be
\begin{eqnarray}
    \Pi (r) = (\Delta \mathcal{A} + \mathcal{C}) r^{\Delta-4} + (4- \Delta) \mathcal{B} r^{-\Delta} + \Delta \mathcal{C} r^{\Delta-4} \ln r
\end{eqnarray}
Since, 
\begin{eqnarray}
    \text{ Im}(\mathcal{A}^*\mathcal{C}) = DB; \quad \text{ Im}(\mathcal{B}^*\mathcal{C}) = \text{ Im}(\mathcal{C}^*\mathcal{B}) = 0; \quad \text{ Im}(\mathcal{C}^*\mathcal{A}) = - DB
\end{eqnarray}
we obtain
\begin{eqnarray}
    \text{ Im}\left[ \phi^*_{\omega, {\bf k}} \Pi \right] \sim (4 - \Delta) (\mathcal{A}^*\mathcal{B})  + \Delta (\mathcal{B}^*\mathcal{A}) 
\end{eqnarray}
as $y \to 0$. We notice that the coefficient of the logarithmic term $\mathcal{C}$ does not enter in the above expression. So, the expression for the ingoing flux near boundary is same as that for the non-integer $\Delta$. This suggests that we can use the same formula to compute greybody factor in the Hankel basis.  

Next we compute the greybody factor to the leading order. As we just mentioned that for computing the greybody factor we shall only consider the normalizable and non-normalizable part of $N_\nu$ since there is no contribution from the logarithmic part in the flux near the asymptotic boundary. From~\eqref{eq:AJ},~\eqref{eq:AN} and~\eqref{eq:BN} we get 
\begin{eqnarray}
A_J (\omega, \mathbf{k})   &=& \frac{4\left( 1 - \frac{\mathbf{k}^2}{\omega^2}\right)^{\frac{n + \epsilon }{2}}}{\Gamma[n+1+\epsilon ]}\left( \frac{\omega}{2}\right)^{n +2+ \epsilon}, \\
A_N (\omega, \mathbf{k}) &=& - \frac{1}{\pi} 2^{-n-\epsilon} \left( 1 - \frac{\mathbf{k}^2}{\omega^2}\right)^{\frac{n + \epsilon }{2}} \omega^{n+2+\epsilon} \cos \left( \pi(n+2+\epsilon) \right) \Gamma[-n-\epsilon]\\
B_N  (\omega, \mathbf{k}) &=& - \frac{\Gamma[n+\epsilon]\,2^{n+\epsilon}}{\pi} \left( 1 - \frac{\mathbf{k}^2}{\omega^2}\right)^{-\frac{n+\epsilon}{2}} \omega^{2-n-\epsilon}
\end{eqnarray}

The connection matrix elements for this case become
\begin{equation}\label{eq:connection_matrix_at}
\begin{array}{|c|c|}
\hline \mathcal{M}_{++}\left(a_h, a ; a_0\right)  &  \frac{\Gamma\left(-2 a \right)}{\Gamma\left(\frac{1}{2}-a + \frac{i r_h \omega}{4}\right)} \frac{\Gamma\left(1+\frac{i r_h \omega}{2}\right)}{\Gamma\left(\frac{1}{2}-a+\frac{i r_h \omega}{4}\right)} \\
\hline \mathcal{M}_{+-}\left(a_h, a ; a_0\right)  &  \frac{\Gamma\left(2 a \right)}{\Gamma\left(\frac{1}{2}+ a + \frac{i r_h \omega}{4}\right)} \frac{\Gamma\left(1+\frac{i r_h \omega}{2}\right)}{\Gamma\left(\frac{1}{2}+a+\frac{i r_h \omega}{4}\right)} \\
\hline \mathcal{M}_{-+}\left(a_h, a ; a_0\right) & \frac{\Gamma\left(-2 a \right)}{\Gamma\left(\frac{1}{2}-a - \frac{i r_h \omega}{4}\right)} \frac{\Gamma\left(1-\frac{i r_h \omega}{2}\right)}{\Gamma\left(\frac{1}{2}-a-\frac{i r_h \omega}{4}\right)} \\
\hline \mathcal{M}_{--}\left(a_h, a ; a_0\right)   & \frac{\Gamma\left(2 a \right)}{\Gamma\left(\frac{1}{2}+ a - \frac{i r_h \omega}{4}\right)} \frac{\Gamma\left(1-\frac{i r_h \omega}{2}\right)}{\Gamma\left(\frac{1}{2}+a-\frac{i r_h \omega}{4}\right)} 
\\
\hline \mathcal{M}_{++}\left(a, a_1 ; a_{\infty}\right) & \frac{\Gamma\left(-n - \epsilon \right)}{\Gamma\left(\frac{1}{2}- \frac{n + \epsilon }{2}+a + \frac{r_h \omega}{4}\right)} \frac{\Gamma\left(1+2 a\right)}{\Gamma\left(1+a-\frac{n + \epsilon }{2} - \frac{ r_h \omega}{4}\right)} \\ 
\hline \mathcal{M}_{+-}\left(a, a_1 ; a_{\infty}\right)  & \frac{\Gamma(n+ \epsilon)}{\Gamma\left(\frac{1}{2}+ \frac{n + \epsilon}{2}+a - \frac{ r_h \omega}{4}\right)} \frac{\Gamma\left(1+2 a\right)}{\Gamma\left(\frac{1}{2}+ \frac{n+ \epsilon}{2}+a +\frac{ r_h \omega}{4}\right)}\\ 
\hline \mathcal{M}_{-+}\left(a, a_1 ; a_{\infty}\right) & \frac{\Gamma\left(-n-\epsilon\right)}{\Gamma\left(\frac{1}{2}- \frac{n+\epsilon}{2}-a + \frac{r_h \omega}{4}\right)} \frac{\Gamma\left(1-2 a\right)}{\Gamma\left(\frac{1}{2}- \frac{n + \epsilon}{2}-a - \frac{ r_h \omega}{4}\right)} \\
\hline \mathcal{M}_{--}\left(a, a_1 ; a_{\infty}\right) & \frac{\Gamma(n + \epsilon)}{\Gamma\left(\frac{1}{2}+ \frac{n + \epsilon }{2}-a - \frac{ r_h \omega}{4}\right)} \frac{\Gamma\left(1-2 a\right)}{\Gamma\left(\frac{1}{2}+ \frac{n + \epsilon}{2}-a +\frac{ r_h \omega}{4}\right)} \\
\hline
\end{array}
\end{equation}
Notice that both $C_{21}$ and $A_N$ diverge as $\epsilon  \to 0$. Since $C_{22}$ is finite in general we find
\begin{eqnarray}
    \frac{\left| i \frac{B_N}{A_{12}} C_{21} - \frac{A_J + i A_N}{A_{11}} C_{22} \right|^2}{\left| i \frac{B_N}{A_{12}} C_{21} + \frac{A_J-  i A_N}{A_{11}} C_{22} \right|^2} \approx \frac{\left| i \frac{B_N}{A_{12}} C_{21} - \frac{i A_N}{A_{11}} C_{22} \right|^2}{\left| i \frac{B_N}{A_{12}} C_{21} - \frac{i A_N}{A_{11}} C_{22}\right|^2} = 1
\end{eqnarray}
where we have neglected the terms $\frac{A_J}{A_{11}}C_{22}$ and $-\frac{A_J}{A_{11}}C_{22}$ from the numerator and the denominator respectively. 

Therefore the greybody factor vanishes in such cases:
\begin{eqnarray}
    \gamma_{\mathbf{k}} (\omega) \approx  1 - \frac{\left| i \frac{B_N}{A_{12}} C_{21} - \frac{i A_N}{A_{11}} C_{22} \right|^2}{\left| i \frac{B_N}{A_{12}} C_{21} - \frac{i A_N}{A_{11}} C_{22}\right|^2} = 1-1 =0
\end{eqnarray}
i.e., for such radiations the asymptotic boundary of the AdS black brane behaves like a perfect reflector. In the next subsection, we explicitly analyze the case where $\Delta=3$ and use numerical analysis to show that the greybody factor drops to zero.
\subsection{Explicit Calculations with $\Delta = 3$}
In this subsection,  we shall compute the greybody factor for $\Delta = 3$ explicitly. The asymptotic field equation~\eqref{eq:scalar_ODE_near_bdy} for this case becomes 
\begin{eqnarray}
\label{eq:delta_3_ode}
\left[r^2 \partial_r^2 - 3 r \partial_r + (\omega^2 - \mathbf{k}^2) r^2 +3  \right] \phi_{\omega,{\bf{k}}}(r)   = 0
\end{eqnarray}
since $m^2 = \Delta (\Delta -4) = - 3$. We consider a  series solution near the regular singular point $r=0$
\begin{equation}
    \phi (r) = \sum_{m=0}^\infty a_m r^{m+\sigma}; \qquad \text{ with } a_0 \neq 0
\end{equation}
using the Frobenius method. The indicial equation reads
\begin{equation}
    \sigma^2 - 4 \sigma + 3  = 0
\end{equation}
with two solutions $ \sigma = 3 $ and $1$. For $\sigma = 3$, we solve the recursion relations of $a_m$ and find the solution to be
\begin{eqnarray}
    \phi_1 (r) = a_0 \frac{2}{\beta} r^2 \, J_1 (\beta r) 
\end{eqnarray}
where $\beta = \sqrt{\omega^2 - \mathbf{k}^2 }$ and $J_1 (\beta r)$ is the Bessel function of the first kind. The solution for $\sigma = 1$ becomes
\begin{eqnarray}
    \phi_2 (r) = \frac{\pi \beta}{4}r^2 N_1 ( \beta r) + \frac{\beta}{2} \left( \frac{3}{4} + \ln 2 - \gamma \right) r^2 J_1 (\beta r)
\end{eqnarray}
where $\gamma$ is the Euler's constant and $N_1 (\beta r)$ is the Bessel function of the second kind (i.e Neumann functions). 

Now from the above solution of the field equation near the asymptotic boundary we cal read the following coefficients:
\begin{eqnarray}
    A_J (\omega, \mathbf{k}) &=& \frac{\omega^2 \beta }{2} \\
    A_N (\omega, \mathbf{k}) &=& \frac{\omega^2 \beta }{\pi} \left(\gamma + \ln \frac{\beta}{2} - \frac{1}{2} \right) \\
    B_N (\omega, \mathbf{k}) &=& -\frac{2\omega^2}{\pi \beta } \\
    C_N (\omega, \mathbf{k}) &=& \frac{\omega^2 \beta }{\pi}; \quad (\text{coefficient of } r^\Delta \ln r)
\end{eqnarray}
Next we consider the connection formula, relating the solutions around horizon to those at the boundary, when written in the Heun basis as follows
\begin{eqnarray} \label{eq:connection_formula1}
    \psi_{\omega,{\bf{k}}}^{(z_h,\text{in})}(z) = C_{21} \psi_{\omega,{\bf{k}}}^{(11)}(z) + C_{22} \psi_{\omega,{\bf{k}}}^{(12)}(z)
\end{eqnarray}
where $C_{21}$ and $C_{22}$ are given in~\eqref{scalar_c21} and~\eqref{scalar_c22} respectively.

The solutions, $\psi_{\omega,{\bf{k}}}^{(11)}(z)$ and $\psi_{\omega,{\bf{k}}}^{(12)}(z)$, near the boundary can be written in terms of HeunG function as
\begin{equation}
       \psi_{\omega,{\bf{k}}}^{(11)}(z)= (1-z)^{\frac{1}{2}+a_1} \text{HeunG}\left(z_h,q+\alpha(\delta-\beta),\alpha,\delta+\gamma-\beta,\delta, \gamma,z_h\frac{1-z}{(z_h-z)} \right)
\end{equation}
and
\begin{eqnarray}
       && \psi_{\omega,{\bf{k}}}^{(12)}(z)= (1-z)^{\frac{1}{2}-a_1} \text{HeunG}\Big(z_h,q-\alpha(\beta+\delta-2)+(\gamma-1)(\alpha+\beta-1-z_h \gamma),  \\ && \quad \quad \quad \quad \quad \quad \quad \quad \quad \quad \quad \quad \quad \gamma+1-\delta,1+\gamma-\beta,2-\delta,\gamma,z_h\frac{1-z}{(z_h-z)} \Big)
\end{eqnarray}
where, 
\begin{equation} \label{par_rel}
\begin{split}
&\alpha = 1 - a_{0} - a_{1} - a_{h} + a_{\infty}, \\ &\beta = 1 - a_{0} - a_{1} - a_{h} - a_{\infty}, \\
&\gamma = 1 - 2a_{0}, \\ &\delta = 1 - 2a_{1},  \\ &\epsilon = 1 - 2a_{h},
\end{split}
\end{equation}
Using above, one can collectively write the connection formula \eqref{eq:connection_formula1} as follows
\begin{equation} \label{connection formula 1}
\begin{split}
   \psi_{\omega,{\bf{k}}}^{(z_h,\text{in})}(z)= & e^{-\frac{1}{2}(\partial_{a_1}+\partial_{a_h})F}\Gamma(-2a_1) (1-z)^{\frac{1}{2}+a_1} (1-t)^{-a_1}A(a_1) \left[ 1+ \frac{\tilde{A}(a_1)}{2(1+2a_1)}(1-z)+...\right] \\
   & + e^{\frac{1}{2}(\partial_{a_1}-\partial_{a_h})F} \Gamma(2a_1) (1-z)^{\frac{1}{2}-a_1} (1-t)^{a_1}B(a_1) \left[ 1+ \frac{\tilde{B}(a_1)}{2(1-2a_1)}(1-z)+...\right]
\end{split}
\end{equation}
Here the functions $A(a_1)$ and $B(a_1)$, being parts of connection coefficients $C_{11}$ and $C_{12}$ respectively, are given by
\begin{equation}
\begin{split}
    & A(a_1) = z_h^{\frac{1}{2}-a_0-a_h} (1-z_h)^{a_h}  
      \\ & \quad \quad  \left[ \frac{\Gamma(-2a)\Gamma(1-2a_h)}{\Gamma(\frac{1}{2}-a_h-a+a_0)\Gamma(\frac{1}{2}-a_h-a-a_0)} \frac{\Gamma(1-2a)}{\Gamma(\frac{1}{2}-a-a_1+a_\infty)\Gamma(\frac{1}{2}-a-a_1-a_\infty)}z_h^ae^{-\frac{1}{2}\partial_aF} \right. \\ 
      & \quad \quad \left. +  \frac{\Gamma(2a)\Gamma(1-2a_h)}{\Gamma(\frac{1}{2}-a_h+a+a_0)\Gamma(\frac{1}{2}-a_h+a-a_0)} \frac{\Gamma(1+2a)}{\Gamma(\frac{1}{2}+a-a_1+a_\infty)\Gamma(\frac{1}{2}+a-a_1-a_\infty)}z_h^{-a}e^{\frac{1}{2}\partial_aF} \right] 
      \\
\end{split}
\end{equation}
and 
\begin{equation}
    \begin{split}
        & B(a_1) = z_h^{\frac{1}{2}-a_0-a_h} (1-z_h)^{a_h} 
      \\ & \quad \quad \left[ \frac{\Gamma(-2a)\Gamma(1-2a_h)}{\Gamma(\frac{1}{2}-a_h-a+a_0)\Gamma(\frac{1}{2}-a_h-a-a_0)} \frac{\Gamma(1-2a)}{\Gamma(\frac{1}{2}-a+a_1+a_\infty)\Gamma(\frac{1}{2}-a+a_1-a_\infty)}z_h^ae^{-\frac{1}{2}\partial_aF} \right. \\ 
      & \quad \quad \left. +  \frac{\Gamma(2a)\Gamma(1-2a_z)}{\Gamma(\frac{1}{2}-a_h+a+a_0)\Gamma(\frac{1}{2}-a_h+a-a_0)} \frac{\Gamma(1+2a)}{\Gamma(\frac{1}{2}-a+a_1+a_\infty)\Gamma(\frac{1}{2}-a+a_1-a_\infty)}z_h^{-a}e^{\frac{1}{2}\partial_aF} \right]
    \end{split}
\end{equation}
where Nekrasov Shatashvili (NS)
partition function \cite{Matone:1995rx,Bonelli:2021uvf,Dodelson:2022yvn} upto first order in $z_h$ is 
\begin{equation}
    F=\frac{(4 a^2 - 4 a_0^2 + 4 a_h^2 - 1)(4 a^2 + 4 a_1^2 - 4 a_\infty^2 - 1)}{8 - 32 a^2}z_h+ \mathcal{O}(z_h^2)
\end{equation}
%
Moreover,
\begin{eqnarray}
    \tilde{A}(a_1)&=&\frac{1}{ (z_h-1)}\left[(2 a_0^2 (z_h-1)+2 a_0 (2 a_1-1) z_h+2 a_1^2 (z_h+1)-2 a_1 (2 a_\infty +z_h+1) \right. \cr
    &&  \left. +2 \left(-a_\infty^2 z_h+a_\infty^2+a_\infty+a_h^2 z_h-z_h u\right)-2 a_h^2+2 u+1\right]
\end{eqnarray} 
and
\begin{eqnarray}
    \tilde{B}(a_1)&=&\frac{1}{ (z_h-1)}\left[2 a_0^2 (z_h-1)-2 a_0 (2 a_1 z_h+z_h)+2 z_h \left(a_1^2+a_1-a_\infty^2+a_h^2-u\right) \right. \cr 
                   && \left. +2 \left(a_1^2+2 a_1 a_\infty+a_1+a_\infty^2\right)+2 \left( a_\infty-2 a_h^2+2 u+1\right)\right]
\end{eqnarray} 
For $\Delta = 3$, $a_1=\frac{1}{2}$. We take the limit $a_0 \to \frac{1}{2}$ while keeping other Heun parameters fixed. 

Note that the term with $2(1-2a_1)$ in the denominator  in eqn  \eqref{connection formula 1} now becomes crucial as this terms diverges at $a_1=\frac{1}{2}$. Accordingly, we can rewrite with eqn \eqref{connection formula 1} as follows 
\begin{equation}
\begin{split}
   \psi_{\omega,{\bf{k}}}^{(z_h,\text{in})}(z)= & e^{-\frac{1}{2}(\partial_{a_1}+\partial_{a_h})F} \Gamma(-2a_1) (1-z)^{\frac{1}{2}+a_1} A(a_1) \left[ 1+...\right] \\
   & +  e^{\frac{1}{2}(\partial_{a_1}-\partial_{a_h})F} \Gamma(2a_1) (1-z)^{\frac{1}{2}-a_1} B(a_1) \left[ 1+ \frac{\tilde{B}(a_1)}{2(1-2a_1)}(z-1)+...\right]
\end{split}
\end{equation}
Further, we re-arrange the terms and take the limit $a_1\to \frac{1}{2}$ as follows\footnote{Here, we use the relation \begin{equation}
    \Gamma(-2a_0)=\frac{\Gamma(2-2a_0)}{(-2a_0)(1-2a_0)}. \nonumber 
\end{equation}}
\begin{equation}
\begin{split}
   \psi_{\omega,{\bf{k}}}^{(z_h,\text{in})}(z)= & \left[ e^{\frac{1}{2}(\partial_{a_1}-\partial_{a_h})F}   B(a_1) \right]\Big|_{a_1=\frac{1}{2}} \\
   & + (1-z) \lim_{a_1\to\frac{1}{2}}\frac{e^{-\frac{1}{2}\partial_{a_h}F}}{(1-2a_1)} \left[
         e^{\frac{1}{2}\partial_{a_1}F} \frac{\Gamma(2a_1)}{2}  B(a_1) \tilde{B}(a_1) (1-z)^{\frac{1}{2}-a_1} \right. \\ & \left.
         \qquad \qquad \qquad   -e^{-\frac{1}{2}\partial_{a_1}F} \frac{\Gamma(2-2a_1)}{(2a_1)}  A(a_1) (1-z)^{-\frac{1}{2}+a_1} 
   \right]
\end{split}
\end{equation}
For the limit to exist, the terms in the second square bracket should vanish at $a_1=\frac{1}{2}$.  This indeed happens because of the following condition
\begin{equation}
    \left[ e^{\partial_{a_1}F} \frac{B(a_1)\tilde{B}(a_1)}{2A(a_1)} \right]\Bigg|_{a_1=\frac{1}{2}}=1.
\end{equation}
After taking the limit, we have
\begin{eqnarray} \label{conn eqn}
  \psi_{\omega,{\bf{k}}}^{(z_h,\text{in})}(z)=  &&\left[  e^{\frac{1}{2}(\partial_{a_1}-\partial_{a_h})F}  B(a_1) \right]\Big|_{a_1=\frac{1}{2}} \cr
   &&+ (1-z) \left[\frac{e^{-\frac{1}{2}(\partial_{a_1}+\partial_{a_h})F}}{2 \sqrt{1-z_h}}\bigg( A'(a_1)+A(a_1)\Big(-2+4\gamma+2\log(1-z)  \right.\cr  && \quad \quad \quad \quad \quad  \left.+\frac{B'(a_1)}{B(a_1)}-F''(a_1)+\frac{4(a_0 z_h-a_\infty)}{(1-z_h)\tilde{B}(a_1)}\Big)\bigg)\right]\Bigg|_{a_1=\frac{1}{2}}
\end{eqnarray}
Here, the superscripts $'$ and $''$ respectively denote the first and second order derivatives with respect to $a_1$. Now, the asymptotic behavior of Heun equation \eqref{heun} for $a_1=\frac{1}{2}$ near the boundary $z=1$  becomes
\begin{equation}
    \Psi(z)= \tilde{p}_1 + \tilde{p}_2  (1-z)
\end{equation}
Finally, from eqn\eqref{conn eqn}, we can identify the connection coefficients\cite{Nunez:2003eq} given by
\begin{eqnarray}
    && C_{21}= \left[\frac{e^{-\frac{1}{2}(\partial_{a_1}+\partial_{a_h})F}}{2 \sqrt{1-z_h}}\bigg( A'(a_1)+A(a_1)\Big(-2+4\gamma  +\frac{B'(a_1)}{B(a_1)}-F''(a_1)+\frac{4(a_0 z_h-a_\infty)}{(1-z_h)\tilde{B}(a_1)}\Big)\bigg)\right]\Bigg|_{a_1=\frac{1}{2}} \notag
    \\ \\ \notag
    \\
    && C_{22}= \left[  e^{\frac{1}{2}(\partial_{a_1}-\partial_{a_h})F}   B(a_1) \right]\Big|_{a_1=\frac{1}{2}}
\end{eqnarray}
where
\begin{equation}
    \begin{split}
        &A'(a_1)|_{a_1=\frac{1}{2}}=  z_h^{\frac{1}{2}-a_0-a_h} (1-z_h)^{a_h}\Gamma(1-2a_h) 
      \\ & \quad \quad  \left[ \frac{\Gamma(-2a)\Gamma(1-2a)\big(\psi(-a - a_\infty)+\psi(-a + a_\infty)-\frac{1}{2}\partial_a\partial_{a_1}F\big)}{\Gamma(\frac{1}{2}-a_h-a+a_0)\Gamma(\frac{1}{2}-a_h-a-a_0)\Gamma(-a+a_\infty)\Gamma(-a-a_\infty)} z_h^ae^{-\frac{1}{2}\partial_aF} \right. \\ 
      & \quad \quad \left. +  \frac{\Gamma(2a)\Gamma(1+2a)\big(\psi(a - a_\infty)+\psi(a + a_\infty)+\frac{1}{2}\partial_a\partial_{a_1}F\big)}{\Gamma(\frac{1}{2}-a_h+a+a_0)\Gamma(\frac{1}{2}-a_h+a-a_0)\Gamma(a+a_\infty)\Gamma(a-a_\infty)} z_h^{-a}e^{\frac{1}{2}\partial_aF} \right] 
    \end{split}
\end{equation}
\begin{equation}
    \begin{split}
        &B'(a_1)|_{a_1=\frac{1}{2}}=-  z_h^{\frac{1}{2}-a_0-a_h} (1-z_h)^{a_h}\Gamma(1-2a_h) 
      \\ & \quad \quad  \left[ \frac{\Gamma(-2a)\Gamma(1-2a)\big(\psi(1-a - a_\infty)+\psi(1-a + a_\infty)+\frac{1}{2}\partial_a\partial_{a_1}F\big)}{\Gamma(\frac{1}{2}-a_h-a+a_0)\Gamma(\frac{1}{2}-a_h-a-a_0)\Gamma(-a+a_\infty)\Gamma(-a-a_\infty)} z_h^ae^{-\frac{1}{2}\partial_aF} \right. \\ 
      & \quad \quad \left. +  \frac{\Gamma(2a)\Gamma(1+2a)\big(\psi(1+a - a_\infty)+\psi(1+a + a_\infty)-\frac{1}{2}\partial_a\partial_{a_1}F\big)}{\Gamma(\frac{1}{2}-a_h+a+a_0)\Gamma(\frac{1}{2}-a_h+a-a_0)\Gamma(a+a_\infty)\Gamma(a-a_\infty)} z_h^{-a}e^{\frac{1}{2}\partial_aF} \right] 
    \end{split}
\end{equation}
and 
\begin{equation}
    \begin{split}
        &\tilde{B}(a_1)|_{a_1=\frac{1}{2}}=  -\frac{1}{2} + 2 a_0^2 - 2 a_\infty^2 + 2 a_h^2 - 2 u
    \end{split}
\end{equation}
in the above, $\psi(z)$ denotes the digamma function i.e. the logarithmic derivative of the Gamma function. Finally, for $r_h = 1$, $\mathbf{k} = 0$, taking the classical block $F$ to the linear order in $z_h$ we numerically compute the greybody factor for $\Delta = 3$ using the above expressions for the connection coefficients and $A_J,\, A_N,\, B_N$ and $C_N$ (all finite) and plot with respect to $\omega$.
\\
\begin{center}
\begin{figure}[h]
    \centering
    \includegraphics[width=0.6\linewidth]{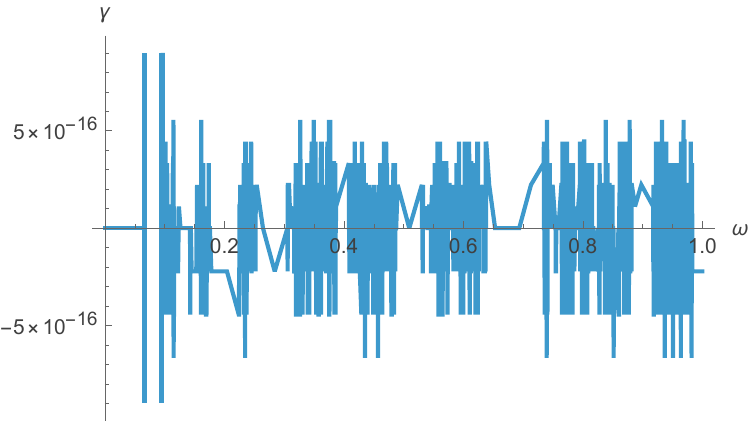}
    \caption{Plot for greybody factor $\gamma$ vs. frequency $\omega$ with $\Delta=3, r_h=1$ and $k=0$. }
    \label{fig:placeholder}
\end{figure}
\end{center}
Note that for the entire range of $\omega$ the value of the greybody factor appears of the order of $\mathcal{O}(10^{-16})$ which can be considered as zero. 
%
%
\section{Maxwell fields in AdS-Schwarzschild black brane} \label{maxwelltheory}
According to AdS/CFT dictionary, Maxwell fields in the bulk contribute to the source term of the boundary $U(1)$ current primary operator. We consider Maxwell fields in the five dimensional AdS Schwarzschild black brane background
\begin{equation}
    S = \int dr \ d^4x \, \sqrt{-g} \frac{1}{4} F^{IJ}F_{IJ}
\end{equation}
where $g$ is the determinant of the metric~\eqref{eq:metric} and
\begin{equation}
    F_{IJ} = \nabla_I A_J - \nabla_J A_I
\end{equation}
The gauge field at the boundary acts as a source term of the current primary operator in the boundary CFT
\begin{equation}
    \label{eq:cft_action}
    S_{CFT} = S_0 + \int d^4x ~\sqrt{-\gamma} a_\mu j^\mu ; 
\end{equation}
where $\gamma$ is the determinant of the induced metric on the conformal boundary of the AdS space and 
\begin{equation}
   \lim_{r \to 0} A_\mu (r, x) = a_\mu (x) \quad \text{(up to a normalization factor)}
\end{equation}

The equation of motion of the bulk gauge field is written as
\begin{equation}
    \label{eq:eom0}
    \nabla_I F^{IJ} = 0 \quad \implies \quad \frac{1}{\sqrt{-g}} \partial_I \left( \sqrt{-g} F^{IJ} \right) = 0
\end{equation}
since $F^{IJ}$ is a totally anti-symmetric tensor.

The action and the equation of motion is invariant under the gauge transformation\footnote{The gauge invariance of the bulk action ensures the conservation of the boundary current operator.}
\begin{equation}
    A_I \to A_I + \partial_I \lambda (r, x)
\end{equation}
However we make the following gauge choice 
\begin{equation}
    \label{eq:gauge_choice}
    A_r = 0
\end{equation}
for fixing the gauge. Still we have residual gauge redundancy which will be fixed by $r$-equation of~\eqref{eq:eom0}
\begin{equation}
     \label{eq:r-equation}
    - \partial_r \partial_t A_t + f(r) \partial_r \partial_i A_i = 0 
\end{equation}
We use~\eqref{eq:r-equation} as a constraint-equation in the following equations 
\begin{eqnarray}
    \label{eq:t-equation}
   && f(r) \partial_r^2 A_t - \frac{f(r)}{r} \partial_r A_t +  \partial^2 A_t -  \partial_t(\partial_i A_i) = 0  \\
    \label{eq:i-equation}
    && f(r) \partial_r^2 A_i -\frac{1}{r}\left(f(r) - r f'(r) \right) \partial_r A_i - \frac{1}{f(r)} \partial_t^2 A_i  + \partial^2 A_i - \frac{1}{f(r)}\partial_i\left(-\partial_t A_t + f(r) \partial_j A_j\right) = 0 \cr 
    &&
\end{eqnarray}
where $\partial_i \partial_i = \partial^2$.

Now we decompose $A_\mu$ as~\cite{Liu:1998ty,DHoker:1998bqu}
\begin{equation}
    \label{eq:decomposition}
    A_\mu = A^\perp_\mu + \partial_\mu \xi  
\end{equation}
where the transverse part of the gauge field is 
\begin{equation}
    A^\perp_\mu = \left( \delta^\nu_\mu - \nabla_\mu \frac{1}{\square} \nabla^\nu \right) A_\nu;  \qquad \square = \nabla^\alpha\nabla_\alpha
\end{equation}
and the parallel component of that
\begin{equation}
    \xi = \frac{1}{\square} \nabla^\mu A_\mu 
\end{equation}
such that 
\begin{equation}
    \label{eq:transverse_condition}
    \nabla^\mu A^\perp_\mu = 0.
\end{equation}
The transverse condition~\eqref{eq:transverse_condition} yields 
\begin{equation}
    -\partial_t A_t^\perp + f(r) \partial_iA_i^\perp = 0
\end{equation}
Now we substitute the decomposition~\eqref{eq:decomposition} in the field equations~\eqref{eq:r-equation},~\eqref{eq:t-equation}, and~ \eqref{eq:i-equation}. The constraint equation becomes 
\begin{eqnarray}
     \label{eq:constraint}
    - \partial_r \partial_t A_t^\perp + f(r) \partial_r \partial_i A_i^\perp &=& 0 \\
    \label{eq:xi-r}
   \square \left( \partial_r  \xi \right)&=& 0
\end{eqnarray}
The $t$-equation becomes
\begin{eqnarray}
    f(r) \partial_r^2 A_t^\perp - \frac{f(r)}{r} \partial_r A_t^\perp +  \partial^2 A_t^\perp -  \partial_t(\partial_i A_i^\perp) &=& 0  \\
    \label{eq:xi-t}
    \partial_r^2 \partial_t \xi -\frac{1}{r} \partial_r \partial_t \xi &=& 0
\end{eqnarray}
The $i$-th equation becomes 
\begin{eqnarray}
    f(r) \partial_r^2 A_i^\perp -\frac{1}{r}\left(f(r) - r f'(r) \right) \partial_r A_i^\perp -  \frac{1}{f(r)}\partial_t^2 A_i^\perp  + \partial^2 A_i^\perp - \frac{1}{f(r)}\partial_i\left(-\partial_t A_t^\perp  + f(r) \partial_j A_j^\perp\right) &=& 0  \cr 
    && \\
    \label{eq:xi-i}
    f(r) \partial_r^2 \partial_i \xi - \frac{1}{r}\left(f(r) - r f'(r) \right) \partial_r \partial_i \xi - \frac{1}{f(r)} (\partial_t^2 - \partial_t) \partial_i \xi &=& 0 \cr 
    && 
\end{eqnarray}
\subsection*{Parallel components}
Let 
\begin{equation}
    \xi (r, x) = \int \frac{d\omega}{2\pi} \int \frac{d^3 {\bf k}}{(2\pi)^3} ~~ \tilde \xi_{\omega} (r, {\bf k}) e^{-i(\omega t + k_i x_i)}
\end{equation}
Substituting the above in~\eqref{eq:xi-r} we get
\begin{eqnarray}
   && \int \frac{d\omega}{2\pi} \int \frac{d^3{\bf k}}{(2\pi)^3} ~~ \left[ \omega^2 - {\bf k}^2 f(r)\right] \left(\partial_r \tilde \xi_\omega (r, {\bf k}) \right) ~~ e^{-i(\omega t + k_i x_i)} = 0 \cr 
   && \cr 
   &\implies& \partial_r \tilde \xi_\omega(r, {\bf k}) = 0 
\end{eqnarray}
i.e., $\tilde \xi$ and hence $\xi$ is independent of $r$. Therefore,~\eqref{eq:xi-t} and~\eqref{eq:xi-i} are trivially satisfied. 

On the other hand, according to AdS/CFT correspondence, we would have
\begin{equation}
    S_{CFT} = S_0 + \int d^4x ~\sqrt{-\gamma}~ a_\mu^\perp j^\mu  + \int d^4x ~\sqrt{-\gamma}~ (\partial_\mu \xi ) j^\mu ;
\end{equation}
where $a_\mu^\perp$ is the boundary value of the transverse components of the gauge field. The last term in the above expression can be written as a total derivative
\begin{equation}
    \int d^4x ~\sqrt{-\gamma}~ \partial_\mu (\xi  j^\mu) \qquad \text{since} \quad \partial_\mu j^\mu = 0
\end{equation}
Therefore, it does not contribute in the partition function and we have
\begin{equation}
    S_{CFT} = S_0 + \int d^4x ~\sqrt{-\gamma}~ a_\mu^\perp j^\mu
\end{equation}
\subsection*{Transverse components}
For the transverse components we use~\eqref{eq:transverse_condition} in~\eqref{eq:constraint} to get the constraint equation for the transverse components
\begin{equation}
    \partial_i A_i^\perp = 0
\end{equation}
Finally, using above we get following field equations for the transverse components of the gauge field\footnote{In the zero-temperature limit ($r_h \to \infty$), these equations match with the corresponding transverse gauge field equations in the pure (Lorentzian) AdS space \cite{Liu:1998ty,DHoker:1998bqu}. }
\begin{eqnarray}
    f(r) \partial_r^2 A_t^\perp - \frac{f(r)}{r} \partial_r A_t^\perp - \frac{1}{f(r)} \partial_t^2 A_t^\perp+  \partial^2 A_t^\perp &=& 0 \\
    f(r) \partial_r^2 A_i^\perp -\frac{1}{r}\left(f(r) - r f'(r) \right) \partial_r A_i^\perp - \frac{1}{f(r)} \partial_t^2 A_i^\perp  + \partial^2 A_i^\perp &=& 0
\end{eqnarray}
Notice that the field equations get decoupled.

Finally we redefine the transverse gauge fields as 
\begin{equation}
    A^\perp_\mu (r, x)  = \frac{1}{r} \bar A^\perp_\mu (r,x)   
\end{equation}
In terms of $\bar A_\mu^\perp$ the gauge field equations read
\begin{eqnarray} 
    f(r) \partial^2_r \bar A_t^\perp - \frac{3f(r)}{r} \partial_r \bar A^\perp_t + \left(- \frac{1}{f(r)} \partial_t^2 + \partial^2 + \frac{3f(r)}{r^2} \right) \bar A_t^\perp &=& 0 \cr \label{ateom}
    && \\
    f(r) \partial_r^2 \bar A^\perp_i - \frac{1}{r} \left( 3f(r)- r f'(r) \right) \partial_r \bar A^\perp_i + \left( - \frac{1}{f(r)}\partial_t^2 + \partial^2 + \frac{3f(r)- r f'(r)}{r^2}\right) \bar A^\perp_i &=& 0 \cr \label{aieom}
    &&
\end{eqnarray}
We observe that the equations that the spatial components of the transverse gauge fields satisfy takes a similar form to those of the scalar fields, with the exception of the mass term; in particular $m^2$ is changed to $-3f(r)+r f'(r)$~\eqref{eom}.\footnote{Notice that in the zero-temperature limit, $r_h \to \infty$, the equations for both temporal and spatial components reduce to that for a scalar field of $m^2 = -3$.}
\subsection{Heun's equation} \label{Heuneq}
As mentioned earlier, in this section we rewrite the transverse field equations as the Heun's equation written in the normal form~\eqref{heun}.
\subsubsection*{Temporal component}
The time and space translation symmetries of the AdS black brane metric motivate us to consider the following mode expansion of $\bar A_t^\perp$ 
\begin{equation} \label{ansatz1}
    \bar A_t^\perp(t,r,{\bf{x}})=\int d\omega \int d^3{\bf{x}} \ e^{-i\omega t} e^{-i {\bf{k}} \cdot {\bf{x}}} \ \mathcal{A}_t(\omega,{\bf{k}};\, r) 
\end{equation}
Substituting the above into \eqref{ateom}, the radial equation for $\bar A_t^\perp$ turns out to be
\begin{equation} \label{ateom1}
   \bigg( f(r) \partial^2_r  - \frac{3f(r)}{r} \partial_r  + \left( \frac{\omega^2}{f(r)}  - {\bf{k}}^2 + \frac{3f(r)}{r^2} \right) \bigg) \mathcal{A}_t(\omega,{\bf{k}};\, r)  = 0
\end{equation}
To recast the above equation \eqref{ateom1} into \eqref{heun}, we make the following change of variable
\begin{equation} \label{eq:heunz}
z =\frac{r_h^2}{r^2+r_{h}^2}
\end{equation}
and redefine the field as
\begin{equation}  \label{trans3}
\mathcal{A}_t(\omega,{\bf{k}};\, r)  =\left(\frac{1}{r^3} \frac{d z}{d r}\right)^{-1 / 2} \mathcal{\bar A}_t(\omega,{\bf{k}};\, z) 
\end{equation}
Then $\mathcal{\bar A}_t(\omega,{\bf{k}};\, z) $ satisfies the Heun's equation~\eqref{heun} with the following parameters
\begin{equation}
     a_0=\frac{1}{2}, \quad \quad  a_1=\frac{1}{2}, \quad \quad  a_h=\frac{1}{4} \sqrt{4-r_h^2 \omega^2},   \quad \quad a_{\infty}=\frac{1}{4} \sqrt{4+r_h^2 \omega^2}
\end{equation} 
and 
\begin{equation}
    u=\frac{r_h^2}{8} (\omega^2-2 {\bf{k}}^2)
\end{equation}
The horizon $r=r_h$ is mapped to $z_h=\frac{1}{2}$, while the origin ($r \to \infty$) and the asymptotic boundary ($r=0$) of the AdS black brane are mapped to $z=0$ and $z=1$ respectively.
\subsubsection*{Spatial components}
Similarly we write $\bar A_i^\perp$ as
\begin{equation} 
    \bar A_i^\perp(t,r,{\bf{x}})=\int d\omega \int d^3{\bf{x}} \ e^{-i\omega t} e^{-i {\bf{k}} \cdot {\bf{x}}} \ \mathcal{A}_i(\omega,{\bf{k}};\, r)  
\end{equation}
Then \eqref{aieom} yields 
\begin{equation} \label{aieom1}
       \left[f(r) \partial_r^2  - \frac{1}{r} \left( 3f(r)- r f'(r) \right) \partial_r  + \left(  \frac{\omega^2}{f(r)} - {\bf{k}}^2 + \frac{3f(r)- r f'(r)}{r^2}\right)  \right]\mathcal{A}_i(\omega,{\bf{k}};\, r) = 0 
\end{equation}
Now we rewrite the above equation in $z$-variable as defined in~\eqref{eq:heunz} and make the following field redefinition
\begin{equation}  \label{trans2}
\mathcal{A}_i(\omega,{\bf{k}};\, r) =\left( \frac{f(r)}{r^3} \frac{d z}{d r}\right)^{-1 / 2} \mathcal{\bar A}_i(\omega,{\bf{k}};\, z) 
\end{equation}
Again~\eqref{aieom1} gets transformed into normal form of Heun's equation \eqref{heun} with the following parameters
\begin{equation} \label{aiheunpar}
    a_0=\frac{1}{2}, \quad \quad  a_1=\frac{1}{2}, \quad \quad  a_h=\frac{i r_h \omega}{4}, \quad \quad a_{\infty}=\frac{r_h \omega}{4}
\end{equation} 
and
\begin{equation}
    u=\frac{r_h^2}{8} (\omega^2-2 {\bf{k}}^2)
\end{equation}
\subsection{Boundary conditions at horizon} \label{bdry cond}
In the following subsections, we study the boundary conditions for both spatial and temporal components of the transverse gauge fields near the horizon $r=r_h$ which is mapped to $z=z_h = \frac{1}{2}$. 
\subsubsection*{Spatial components} 
In the tortoise coordinates 
\begin{equation}
    r_*=\int\frac{1}{f(r)} dr
\end{equation}
near the horizon $r=r_h$ the equations \eqref{aieom1} for $\bar A_i^\perp$ take the following form
\begin{equation} \label{eq:spatial_tortoise}
   \bigg(  \partial^2_{r_*}    +  \omega^2  \bigg) \bar A_i^\perp(r_*) = 0
\end{equation}
There are two kinds of solutions to the above equation
\begin{eqnarray}
    \bar A_i^\perp(r_*) &\sim& e^{- i \omega r_*} \qquad \text{(in-going)} \\
                        &\sim& e^{ i \omega r_*} \qquad ~~\, \text{(out-going)}
\end{eqnarray}
We recast the above solutions in the $r$-coordinate and write the general solution as a superposition of those 
\begin{equation}
  \left.  \bar A_i^\perp(r) \right|_{r=r_h}= C_1 \times (r-r_h)^{-\frac{i \omega}{f'(r_h)}} (1+ \cdots)+ C_2 \times (r-r_h)^{\frac{i \omega}{f'(r_h)}} (1 + \cdots )
\end{equation}
where $C_1$ and $C_2$ are integration constants and $f'(r) = df/dr$. We now impose the ingoing boundary condition at horizon 
by setting $C_2=0$ and then we have
\begin{equation} \label{aperphorizon}
  \left.  \bar A_i^\perp(r) \right|_{r=r_h}= C_1 \times (r-r_h)^{-\frac{i \omega}{f'(r_h)}} (1 + \cdots) = C_1 \times (r-r_h)^{\frac{i \omega r_h}{4}} (1 + \cdots )
\end{equation} 
This is the desired behaviour of $\bar A_i^\perp$ near horizon. 

Now the solution of the Heun's equation \eqref{heun} around $z=z_h$ leads to
\begin{equation} \label{aizhor}
 \left. \mathcal{\bar A}_i(\omega,{\bf{k}};\, z)  \right|_{z=z_h} = A (z_h-z)^{\frac{1}{2}+a_h}(1 +\dots ) + B (z_h-z)^{\frac{1}{2}-a_h}(1 +\dots )
\end{equation}
Substituting the expression~\eqref{aiheunpar} for $a_h$ and $z_h=1/2$ obtained in the previous section in the above solution and considering the change of variable~\eqref{eq:heunz} and field redefinition~\eqref{trans2} we recognize the in-going solution in the $z$-coordinate to be
\begin{equation} 
 \left. \mathcal{\bar A}_i(\omega,{\bf{k}};\, z)  \right|_{z=z_h} \sim (z_h-z)^{\frac{1}{2}-a_h} 
\end{equation}
\subsubsection*{Temporal component} 
In contrast to the spatial components of the transverse gauge fields, in the near horizon limit, the equation~\eqref{ateom1} for the temporal component takes the following form when written in the tortoise coordinates
\begin{equation} \label{ateom2}
   \bigg(  \partial^2_{r_*}  + \frac{4}{r_h} \partial_{r_*}  +  \omega^2  \bigg) \bar A_t^\perp(r_*) = 0
\end{equation}
If $r_h$ is very large, i.e., $r_h\gg 4$, the above equation turns to be of the similar form as in~\eqref{eq:spatial_tortoise}. Then we write the in-going solution for the temporal component of the transverse gauge field around $z=z_h$ as
\begin{equation} \label{eq:at_ingoing}
 \left. \mathcal{\bar A}_t(\omega,{\bf{k}};\, z)  \right|_{z=z_h} \sim (z_h-z)^{\frac{1}{2}-a_h} 
\end{equation}
where in the limit $r_h\gg 4$ the Heun's parameters $a_h$ and $a_\infty$ are given by
\begin{equation}
     a_h=\frac{i r_h \omega}{4},   \quad \quad a_{\infty}=\frac{r_h \omega}{4} 
\end{equation} 
which are same as those for the spatial components. So, if $r_h$ is considered to be very large, both the temporal and spatial components of the transverse gauge fields behave in a similar way.

However, when $r_h$ is not so large, the temporal component $\bar A_t^\perp$ satisfies \eqref{ateom2} near the horizon. We identify~\eqref{ateom2} to be the wave equation in an absorbing medium with an attenuation constant $b = 2/r_h$. In this scenario, there are three possibilities:
\begin{enumerate}
    \item \underline{$b^2 < \omega^2$}: In this case, the solution of \eqref{ateom2} is given by
    \begin{equation}
         \bar A_t^\perp(r_*)=e^{-\frac{2r_*}{r_h}} \big( C_1 e^{-i s r_*}+ C_2 e^{i s r_*} \big)
    \end{equation}
    where $s=\sqrt{\omega^2-b^2}$ and $s>0$. The ingoing boundary condition at the horizon claims that $C_2=0$. Further, in the $z$-coordinate the in-going solution near $z=z_h$ will take the same form as in~\eqref{eq:at_ingoing}, with the Heun's parameter
    \begin{equation}
        a_h = \frac{i r_h s}{4} = \frac{i}{4} \sqrt{rh^2 \omega^2 - 4}
    \end{equation}
    \item \underline{$b^2 > \omega^2$}: In this case, the solution of \eqref{ateom2} is given by
     \begin{equation}
         \bar A_t^\perp(r_*)=e^{-\frac{2r_*}{r_h}} \big( C_1 e^{ \bar{s} r_*}+ C_2 e^{- \bar{s} r_*} \big)
    \end{equation}
    where $\bar{s}=\sqrt{b^2-\omega^2}$ and $ \bar s < 2/r_h$. In this case near the horizon ($r_* \to \infty$ at horizon), we do not get wavelike solution and the field exponentially vanishes at the horizon. However, the Heun's parameter is real 
    \begin{eqnarray}
         a_h = \frac{ r_h \bar s}{4} = \frac{1}{4} \sqrt{4- rh^2 \omega^2}
    \end{eqnarray}
    \item \underline{$b^2 = \omega^2$}: In this case, near the horizon $\bar A_t^\perp$ mode behaves as
    \begin{equation}
        \bar A_t^\perp(r_*)=\big( C_1 + C_2  r_* \big) e^{-r_* \omega} 
    \end{equation}
    Again, there is no wavelike solution and the field exponentially vanishes as approaching to the horizon, but the decay is faster as compared to the previous case. In this case, $a_h=0$.
\end{enumerate}
\subsection{Asymptotic behaviour} \label{asymptotic}
Near the asymptotic boundary ($r\to 0$) of the AdS black brane, we observe that the temporal and spatial components of the transverse gauge field satisfy the same equation, in contrast to the horizon
\begin{eqnarray}
\left[r^2 \partial_r^2 - 3 r \partial_r + (\omega^2 - \mathbf{k}^2) r^2 - (-3) \right] \mathcal{A}_\mu (\omega,{\bf{k}};\, r)   = 0
\end{eqnarray}
We write the equation in $u = \omega r$ variable as
\begin{eqnarray}
\left[u^2 \partial_u^2 - 3 u\partial_u + \left(1 - \frac{\mathbf{k}^2}{\omega^2} \right) u^2 - (-3) \right] \mathcal{A}_\mu(\omega,{\bf{k}};\, r)   = 0
\end{eqnarray}
We identify the above equation as Bessel's equation. The linearly independent solutions can be written in terms of the Bessel (J) and Neumann (N) functions. The asymptotic behaviours near $u=0$ are as follows 
\begin{eqnarray}
u^2 J_1 \left( \sqrt{ 1 - \frac{\mathbf{k}^2}{\omega^2} } u\right) &\simeq& \frac{1}{2} \left( 1 - \frac{\mathbf{k}^2}{\omega^2}\right)^{1/2} u^3 = A_J (\omega, \mathbf{k}) r^3 \\
u^2 N_1 \left( \sqrt{ 1 - \frac{\mathbf{k}^2}{\omega^2} } u\right) &\simeq& - \frac{2}{\pi} \left( 1 - \frac{\mathbf{k}^2}{\omega^2}\right)^{-1/2} u + \frac{\Gamma[-1]}{2\pi} \left( 1 - \frac{\mathbf{k}^2}{\omega^2}\right)^{1/2} u^3 \cr 
&=& A_N (\omega, \mathbf{k})  r^3 + B_N (\omega, \mathbf{k})  r
\end{eqnarray}
where
\begin{eqnarray}
A_J (\omega, \mathbf{k})   &=& \frac{\omega^3}{2} \left( 1 - \frac{\mathbf{k}^2}{\omega^2}\right)^{-1/2}, \\
A_N (\omega, \mathbf{k}) &=&  \frac{\Gamma[-1 - \epsilon]}{2\pi} \left( 1 - \frac{\mathbf{k}^2}{\omega^2}\right)^{1/2} \\
B_N  (\omega, \mathbf{k}) &=& - \frac{2}{\pi} \left( 1 - \frac{\mathbf{k}^2}{\omega^2}\right)^{-1/2}
\end{eqnarray}
The solution, that behaves as $\simeq r^3 (\simeq r)$ near the asymptotic boundary, is identified as normalizable (non-normalizable) mode.

On the other hand, near $z=1$ which corresponds to the boundary limit in $z$-coordinate, the linearly independent solutions of the Heun's equation behave as follows
\begin{eqnarray}
     \mathcal{\bar A}_\mu^{(1,+)}(\omega,{\bf{k}};\, z) \simeq (1-z)^{\frac{1}{2} + a_1} &\simeq& r_h^{-2} \, r^2 \\
     \mathcal{\bar A}_\mu^{(1,-)}(\omega,{\bf{k}};\, z) \simeq (1-z)^{\frac{1}{2} - a_1} &\simeq& 1
\end{eqnarray}
since $a_1 = 1/2$ for $ \mathcal{\bar A}_\mu (\omega,{\bf{k}};\, z) $. Using~\eqref{trans3} and~\eqref{trans2}, near the asymptotic boundary we find
\begin{eqnarray}
    \mathcal{A}_\mu(\omega,{\bf{k}};\, r) \simeq \left( A_{1-}  \, r + \cdots \right) +  \left( A_{1+} \, r^3 + \cdots \right)
\end{eqnarray}
where 
\begin{eqnarray}
    A_{1-} = - \frac{i r_h}{\sqrt{2}}, \qquad \text{and} \qquad A_{1+} = - \frac{i}{\sqrt{2} r_h}
\end{eqnarray}
\subsection{Connection formulae}
The linearly independent solutions of the Heun's equation~\eqref{heun} around each singularity are related to those around the other via connection coefficients. Specifically we relate the in-going solution $\mathcal{\bar A}_\mu^{(0,\text{in})}$ near $z=z_0$ to the solutions $\mathcal{\bar A}_\mu^{(1,\pm)}$ near $z=1$ as~\cite{Bonelli:2022ten,Dodelson:2022yvn}
\begin{eqnarray} \label{eq:connection_formula}
    \mathcal{\bar A}_\mu^{(0,\text{in})}(\omega,{\bf{k}};\, z) = C^{\text{(in)}}_+ \mathcal{\bar A}_\mu^{(1,+)}(\omega,{\bf{k}};\, z) + C^{\text{(in)}}_- \mathcal{\bar A}_\mu^{(1,-)}(\omega,{\bf{k}};\, z)
\end{eqnarray}
where
\begin{eqnarray} \label{eq:connection_coefficient}
    C^{\text{(in)}}_{\theta} = z_h^{\frac{1}{2}-a_0-a_h} (1-z_h)^{a_h - a_1} e^{-\frac{1}{2}(\partial_{a_h} + \theta \partial_{a_1})F} \sum_{\theta' = \pm} \mathcal{M}_{-\theta'} (a_h, a; a_0) \mathcal{M}_{(-\theta')\theta} (a,a_1; a_\infty) z_h^{\theta' a} e^{-\frac{\theta'}{2}\partial_a F}
\end{eqnarray}
with
\begin{eqnarray} \label{eq:matrix_elements}
    \mathcal{M}_{\theta\theta'} (a_0, a_1; a_2) = \frac{\Gamma(-2 \theta' a_1)}{\Gamma\left( \frac{1}{2} +\theta a_0 - \theta' a_1 + a_2\right)} \frac{\Gamma(1+ 2 \theta a_0)}{\Gamma\left( \frac{1}{2} +\theta a_0 - \theta' a_1 - a_2\right)}
\end{eqnarray}

For example, the connection matrix for the spatial components of the transverse gauge fields is given by
\begin{equation}\label{eq:connection_matrix_at}
\begin{array}{|c|c|}
\hline \mathcal{M}_{++}\left(a_h, a ; a_0\right)  &  \frac{\Gamma\left(-2 a \right)}{\Gamma\left(\frac{1}{2}-a + \frac{i r_h \omega}{4}\right)} \frac{\Gamma\left(1+\frac{i r_h \omega}{2}\right)}{\Gamma\left(\frac{1}{2}-a+\frac{i r_h \omega}{4}\right)} \\
\hline \mathcal{M}_{+-}\left(a_h, a ; a_0\right)  &  \frac{\Gamma\left(2 a \right)}{\Gamma\left(\frac{1}{2}+ a + \frac{i r_h \omega}{4}\right)} \frac{\Gamma\left(1+\frac{i r_h \omega}{2}\right)}{\Gamma\left(\frac{1}{2}+a+\frac{i r_h \omega}{4}\right)} \\
\hline \mathcal{M}_{-+}\left(a_h, a ; a_0\right) & \frac{\Gamma\left(-2 a \right)}{\Gamma\left(\frac{1}{2}-a - \frac{i r_h \omega}{4}\right)} \frac{\Gamma\left(1-\frac{i r_h \omega}{2}\right)}{\Gamma\left(\frac{1}{2}-a-\frac{i r_h \omega}{4}\right)} \\
\hline \mathcal{M}_{--}\left(a_h, a ; a_0\right)   & \frac{\Gamma\left(2 a \right)}{\Gamma\left(\frac{1}{2}+ a - \frac{i r_h \omega}{4}\right)} \frac{\Gamma\left(1-\frac{i r_h \omega}{2}\right)}{\Gamma\left(\frac{1}{2}+a-\frac{i r_h \omega}{4}\right)} 
\\
\hline \mathcal{M}_{++}\left(a, a_1 ; a_{\infty}\right) & \frac{\Gamma\left(-1 - \epsilon \right)}{\Gamma\left(\frac{1}{2}- \frac{1 + \epsilon }{2}+a + \frac{r_h \omega}{4}\right)} \frac{\Gamma\left(1+2 a\right)}{\Gamma\left(1+a-\frac{1 + \epsilon }{2} - \frac{ r_h \omega}{4}\right)} \\ 
\hline \mathcal{M}_{+-}\left(a, a_1 ; a_{\infty}\right)  & \frac{\Gamma(1+ \epsilon)}{\Gamma\left(\frac{1}{2}+ \frac{1 + \epsilon}{2}+a - \frac{ r_h \omega}{4}\right)} \frac{\Gamma\left(1+2 a\right)}{\Gamma\left(\frac{1}{2}+ \frac{1+ \epsilon}{2}+a +\frac{ r_h \omega}{4}\right)}\\ 
\hline \mathcal{M}_{-+}\left(a, a_1 ; a_{\infty}\right) & \frac{\Gamma\left(-1-\epsilon\right)}{\Gamma\left(\frac{1}{2}- \frac{1+\epsilon}{2}-a + \frac{r_h \omega}{4}\right)} \frac{\Gamma\left(1-2 a\right)}{\Gamma\left(\frac{1}{2}- \frac{1 + \epsilon}{2}-a - \frac{ r_h \omega}{4}\right)} \\
\hline \mathcal{M}_{--}\left(a, a_1 ; a_{\infty}\right) & \frac{\Gamma(1 + \epsilon)}{\Gamma\left(\frac{1}{2}+ \frac{1 + \epsilon }{2}-a - \frac{ r_h \omega}{4}\right)} \frac{\Gamma\left(1-2 a\right)}{\Gamma\left(\frac{1}{2}+ \frac{1 + \epsilon}{2}-a +\frac{ r_h \omega}{4}\right)} \\
\hline
\end{array}
\end{equation}
In the large $r_h$ limit, the connection matrix for the temporal component becomes same as above.
\subsection{Greybody factor}
We saw in section~\ref{asymptotic} that the field equations for both the temporal and spatial components of the transverse gauge fields near the asymptotic boundary of the AdS Schwarzschild black brane take the similar form as those for a massive scalar with $m^2 = -3$. Since $m^2 \neq 0$, we can use the Heun functions to find the greybody factor of the black brane as discussed in~\cite{Noda:2022zgk}. 

First we construct IN mode of the transverse gauge fields as 
\begin{eqnarray} \label{eq:in_mode_1}
   \mathcal{\bar A}_\mu^{\text{IN}}(\omega,{\bf{k}};\, z)  &=& \left\{ \begin{matrix} \mathcal{\bar A}_\mu^{(z_h,\text{in})}(\omega,{\bf{k}};\, z) \simeq (z_h-z)^{\frac{1}{2} -a_h} \qquad \qquad \qquad ~z \to z_h\\ \\ C^{\text{(in)}}_+ \mathcal{\bar A}_\mu^{(1,+)}(\omega,{\bf{k}};\, z) + C^{\text{(in)}}_- \mathcal{\bar A}_\mu^{(1,-)}(\omega,{\bf{k}};\, z)  \qquad z \to 1 \end{matrix} \right. 
\end{eqnarray}
ggNow, $\mathcal{\bar A}_\mu^{(1,+)}$ and $\mathcal{\bar A}_\mu^{(1,-)}$ behaves as $\sim r^2$ and $\sim 1$ respectively near the asymptotic boundary. Since we need in/out-going modes at the asymptotic boundary to compute the greybody factor, we change the basis of the IN mode by defining 
\begin{eqnarray}
    \mathcal{\bar A}_\mu^{+} (\omega,{\bf{k}};\, z) &=& \frac{A_J + i A_N}{A_{1+}} \mathcal{\bar A}_\mu^{(1,+)}(\omega,{\bf{k}};\, z) + i \frac{B_N}{A_{1-}} \mathcal{\bar A}_\mu^{(1,-)}(\omega,{\bf{k}};\, z) \\
    \mathcal{\bar A}_\mu^{-} (\omega,{\bf{k}};\, z) &=& \frac{A_J - i A_N}{A_{1+}} \mathcal{\bar A}_\mu^{(1,+)}(\omega,{\bf{k}};\, z) - i \frac{B_N}{A_{1-}} \mathcal{\bar A}_\mu^{(1,-)}(\omega,{\bf{k}};\, z)
\end{eqnarray}
such that the IN modes become
\begin{eqnarray} \label{eq:in_mode_2}
   \mathcal{\bar A}_\mu^{\text{IN}}(\omega,{\bf{k}};\, z)  &=& \left\{ \begin{matrix} \mathcal{\bar A}_\mu^{(z_h,\text{in})}(\omega,{\bf{k}};\, z) \simeq (z_h-z)^{\frac{1}{2} -a_h} \qquad \qquad \qquad ~z \to z_h\\ \\ C_{\text{out}} \mathcal{\bar A}_\mu^{+}(\omega,{\bf{k}};\, z) + C_{\text{in}} \mathcal{\bar A}_\mu^{-}(\omega,{\bf{k}};\, z)  \qquad \qquad \quad z \to 1 \end{matrix} \right. 
\end{eqnarray}
The asymptotic forms of $\mathcal{\bar A}_\mu^{+}$ and $\mathcal{\bar A}_\mu^{-}$ are the outgoing and ingoing modes respectively. 

Comparing~\eqref{eq:in_mode_1} and~\eqref{eq:in_mode_2} we obtain
\begin{eqnarray}
    C_{\text{out}} &=& \frac{A_{1+}A_{1-}}{2i A_J B_N} \left[ i \frac{B_N}{A_{1-}} C^{\text{(in)}}_+ - \frac{A_J+ i A_N}{A_{1+}} C^{\text{(in)}}_- \right] \\
    C_{\text{in}} &=& \frac{A_{1+}A_{1-}}{2i A_J B_N} \left[ i \frac{B_N}{A_{1-}} C^{\text{(in)}}_+ + \frac{A_J - i A_N}{A_{1+}} C^{\text{(in)}}_- \right]
\end{eqnarray}
Finally we obtain the greybody factor in terms of the connection coefficients of the Heun functions as
\begin{eqnarray}
    \gamma_{\mathbf{k}} (\omega) &=& 1 - \left| \frac{C_{\text{out}}}{C_{\text{in}}} \right|^2 \cr 
                                 &=& 1 -  \frac{\left| i \frac{B_N}{A_{1-}} C^{\text{(in)}}_+ - \frac{A_J + i A_N}{A_{1+}} C^{\text{(in)}}_- \right|^2}{\left| i \frac{B_N}{A_{1-}} C^{\text{(in)}}_+ + \frac{A_J- i A_N}{A_{1+}} C^{\text{(in)}}_-\right|^2}
\end{eqnarray}
Since both $C^{\text{(in)}}_+$ and $A_N$ have the factor $\Gamma(-1)$, the greybody factor vanishes
\begin{eqnarray}
    \gamma_{\mathbf{k}} (\omega) = 1 - \frac{\left| i \frac{B_N}{A_{1-}} C^{\text{(in)}}_+ - \frac{i A_N}{A_{1+}} C^{\text{(in)}}_- \right|^2}{\left| i \frac{B_N}{A_{1-}} C^{\text{(in)}}_+ - \frac{i A_N}{A_{1+}} C^{\text{(in)}}_-\right|^2} = 1-1 =0
\end{eqnarray}
Thus we find that the photon radiations get fully reflected back from the asymptotic boundary of the AdS black brane. Note that this is equivalent to the case $\Delta = 3$ for scalar perturbations, which explicitly shown earlier with a plot of the greybody factor vs. $\omega$ in the $\mathbf{k} = 0$ limit.  

We should mention here that due to the appearance of the poles of the gamma functions in the connection formulae, one expects divergence in the expression of the thermal correlators of the CFT living at the conformal boundary of the AdS black brane. However, one can regularize the correlators (for example, via dimensional regularization) and using appropriate renormalization scheme one can obtain the finite answer. This is elucidated in the Appendix~\ref{correlator}. 
\section{Conclusion} \label{conclusion}
In this work, we computed greybody factor for the Schwarzschild black brane in the five-dimensional AdS spacetime by considering a massive scalar field and the Maxwell field propagating through the black brane background. In particular, first we mapped the linearized equations satisfied by the scalar field and the transverse components of the Maxwell fields to the normal form of Heun differential equation by making a suitable choice of coordinate and field transformations. Notice that the Heun's equation has for regular singularities, two of which correspond to the horizon and the asymptotic boundary of the AdS black brane. Further, we have imposed ingoing boundary condition to the solutions of the field equation near the horizon. At the asymptotic boundary of the AdS black brane, the solutions of the linearized field equations can be written as the linear combination of normalizable modes and non-normalizable modes. We construct IN modes by considering appropriate linear combinations of the normalizable and non-normalizable modes such that the modes behave as ingoing and outgoing waves at the asymptotic boundary. These solutions become the Hankel function of first and second kind respectively in the far-field approximation. However, the solutions near the horizon are related to those near the asymptotic boundary of the AdS black brane via the connections coefficients among the solutions of the Heun's equation near the corresponding singularities. Thus the coefficients of the ingoing and outgoing modes of the asymptotic solutions are obtained in terms of the connection coefficients of the Heun functions. Finally, we compute the greybody factor by calculating the ratio of the coefficient of the outgoing mode of the asymptotic solution to that of the ingoing mode.

We found that the greybody factor for the scalar fields, dual to the conformal scalar primary operator with scaling dimensions to be integers, vanishes. Similarly, the greybody factor for the transverse components of the Maxwell fields also vanishes. Specifically, this happens due to the appearance of the gamma function with negative integer argument in the expression of the connection coefficients. Particularly, for the transverse components of the Mawell fields the factor becomes $\Gamma(-1)$. Physically it means that, for the transverse Maxwell fields and the particular scalar fields mentioned above the AdS boundary behaves as a perfect reflector, i.e., the corresponding emitted radiations from the black brane cannot overcome the potential barrier caused by the spacetime geometry of its surroundings and gets reabsorbed. The vanishing of the greybody factor observed here is rooted in the analytic structure of the connection coefficients of the underlying Heun equation. Similar analytic features have recently been identified in the study of Love numbers of compact objects, where poles arise in the tidal response coefficients~\cite{DiRusso:2024hmd}. Here poles in Love numbers correspond to resonant or enhanced tidal responses.

In this work we have studied the spectroscopy of Schwarzschild AdS black brane with the scalar fields and the Maxwell fields as the probes. The next immediate question one can ask is whether gravitational perturbations also cause the greybody to vanish. Apparently, we can observe that near the asymptotic boundary the linearized equation satisfied by the transverse traceless components of the gravitational modes coincides with the equation for massless scalar field~\cite{Arutyunov:1998ve}. Therefore, the Heun/gravity technique may not be applicable to compute the greybody factor for the gravitons, since the technique developed in~\cite{Noda:2022zgk} is applicable for the massive scalars. However, one can compute the greybody factor for the gravitational fields directly in the far-field approximation. It will be reported soon. 

It will also be very interesting to explore more in this direction to fully understand the physical origin and implications of the vanishing of greybody factor in the context of AdS/CFT correspondence. 

\section*{Acknowledgement}
We thank Shankhadeep Chakrabortty and Julián Barragán Amado for useful discussions. We also thank Hayato Motohashi and Sousuke Noda for their valuable comments. AM would like to thank the Council of Scientific and Industrial Research (CSIR), Government of India, for the financial support through a research fellowship (File No.: 09/1005(0034)/2020-EMR-I). 

\appendix

\section{Classical conformal block} \label{classical_block}
In Liouville CFT, there exist primary operators which have null descendants. Such primary operators are called the degenerate operators. The simplest example of a null descendant in level two of the Virasoro module is
\begin{eqnarray}
    \left( L_{-1}^2 + b^2 L_{-2} \right) | h_{1,2} \rangle  
\end{eqnarray}
where the parameter $b$ is related to the central charge $c$ of the CFT as
\begin{eqnarray}
    c = 1 + 6 \left( b + \frac{1}{b}\right)^2
\end{eqnarray}
and the conformal weight of the degenerate primary operator $\Phi_{h_{1,2}}$ is $h_{1,2} = - \frac{1}{2} - \frac{3}{4b^2}$.

The following five point function with one degenerate operator insertion at $z$
\begin{eqnarray}
    \mathcal{F} (z) = \langle \mathcal{O}_{\Delta_\infty} (\infty) \mathcal{O}_{\Delta_1} (1) \mathcal{O}_{\Delta_h} (z_h) \Phi_{\Delta_{1,2}} (z) \mathcal{O}_{\Delta_0} (0) \rangle 
\end{eqnarray}
satisfies the BPZ equation
\begin{eqnarray}
    && \left( b^{-2}\partial_z^2 + \frac{\Delta_1}{(z-1)^2}- \frac{\Delta_1+z_h \partial_{z_h} + \Delta_h+z \partial_z + \Delta_{2,1}+\Delta_0-\Delta_{\infty}}{z(z-1)} \right. \cr 
    && \qquad \qquad \qquad \qquad \left. + \frac{\Delta_h}{(z-z_h)^2}+ \frac{z_h}{z(z-z_h)}\partial_{z_h} - \frac{1}{z}\partial_z + \frac{\Delta_0}{z^2}\right) \mathcal{F}(z) = 0
\end{eqnarray}
where $\Delta_{\infty},\, \Delta_1,\, \Delta_h$ and $\Delta_0$ are the scaling dimensions of the primary operators inserted at $z = \infty,\, 1,\, z_h$ and $0$ respectively. The BPZ equation has four regular singularities at $z = 0,\, z_h ,\, 1$ and $\infty$.

Now we take the semi-classical limit
\begin{eqnarray}
    b \to 0 \quad \text{ and } \quad \alpha_i \to 0 \quad \text{ such that } \quad b \alpha_i = a_i \text{ (finite) }
\end{eqnarray}
where
\begin{eqnarray}
    \Delta_i = \alpha_i (Q-\alpha_i); \qquad Q = \left( b + \frac{1}{b}\right)
\end{eqnarray}
in which the BPZ equation becomes the Heun's equation.

The five-point function $\mathcal{F} (z)$ obeys the partial wave expansions
\begin{eqnarray}
    \mathcal{F}(z) =\sum_{\Delta,\,\ell} \sum_{\Delta',\,\ell'} \sum_a \lambda_{\Delta_{\infty}\Delta_1(\Delta, \ell)}\lambda^a_{(\Delta,\ell)\Delta_h(\Delta',\ell')} \lambda_{(\Delta',\ell')\Delta_{1,2}\Delta_0} W^{(a)}_{(\Delta,\ell) (\Delta',\ell')}(z)
\end{eqnarray}
with the partial waves
\begin{eqnarray}
    W^{(a)}_{(\Delta,\ell) (\Delta',\ell')}(z) = P(z)\, G^{(a)}_{(\Delta,\ell) (\Delta',\ell')}(z)
\end{eqnarray}
where $\lambda_{\Delta_{\infty}\Delta_1(\Delta, \ell)}$ and $\lambda_{(\Delta',\ell')\Delta_{1,2}\Delta_0}$ are the OPE coefficients, $\lambda^a_{(\Delta,\ell)\Delta_t(\Delta',\ell')}$ is an independent coefficient, and $G^{(a)}_{(\Delta,\ell) (\Delta',\ell')}(z)$ is the five-point conformal block. Here the sum over $a$ denotes the sum over conformal families. In the semi-classical limit, the conformal blocks satisfy the Heun's equation written in the normal form. Thus the solutions around different singularities of the Heun's equation correspond to the semi-classical conformal blocks that appear in the partial wave of expansion of the five-point function in different channels. Since the Heun's equation is a second order ODE, around each singularities we shall have two linearly independent solutions which correspond the semi-classical conformal blocks of two different fusion channels in a given expansion of the correlation function. 

Now, the correlation function admits crossing symmetry, the conformal blocks in different channels are related to each other via crossing relations. These crossing relations give rise to connection formulae among the solutions of Heun's equation around different singularities~\cite{Bonelli:2022ten}. In particular, the solution that goes as
\begin{eqnarray}
    \chi^{z_h, \text{in}} (z) \simeq (z_h - z)^{\frac{1}{2} - a_h}
\end{eqnarray}
around $z =z_h$ is related to the solutions $\chi^{1, \pm}$ around $z=1$ as~\cite{Dodelson:2022yvn}
\begin{eqnarray}
\label{eq:connection}
\chi^{z_h, \text{in}}(z) &=& \sum_{\theta'=\pm} \left(\sum_{\theta=\pm} \mathcal{M}_{-\theta}(a_{t}, a; a_0)\mathcal{M}_{(-\theta)\theta'}(a, a_1; a_{\infty}) \right. \cr  
&& \qquad \qquad \left. t^{\theta a}e^{-\frac{\theta}{2}\partial_a F} \right) t^{\frac{1}{2}-a_0-a_{t}}(1-t)^{a_t-a_1} e^{-\frac{1}{2}(\partial_{a_t}+\theta' \partial_{a_1})F} \chi^{1, \theta'}(z)\nonumber, \\
\end{eqnarray}
where $F$ is four point classical conformal block in Liouville CFT and the connection matrix elements are given in~\eqref{eq:matrix_elements}.
\section{Current-current correlators} \label{correlator}
According to the prescription given in~\cite{Son:2002sd}, to compute the retarded Green functions of the $U(1)$ global current primary operators in the CFT residing at the boundary of the AdS Schwarzschild black brane we first define the source $S_\mu$ and the response $R_\mu$ functions in the asymptotic behaviour of the transverse gauge fields as
\begin{eqnarray} \label{eq:source_response}
    \left. \mathcal{A}_\mu(\omega,{\bf{k}};\, r) \right|_{r \to 0} \simeq S_\mu (\omega, \mathbf{k}) \left( r + \cdots \right) + R_\mu (\omega, \mathbf{k}) \left( r^3 + \cdots \right)
\end{eqnarray}
Then we find the ratio of the response to the source functions which yields the desired Green's function.
Finally relating~\eqref{eq:source_response} to the connection formula~\eqref{eq:connection_formula} we obtain the retarded Green's function of the conserved currents as
\begin{eqnarray}
    \left \langle J_\mu (-\omega, -\mathbf{k}) J_\nu (\omega, \mathbf{k}) \right \rangle_R &=& \delta_{\mu\nu} ~\frac{R_\mu(\omega, \mathbf{k})}{S_\mu(\omega, \mathbf{k})}  
    = \delta_{\mu\nu} \left[r_h^{2a_1} \frac{C^{\text{(in)}}_+ }{C^{\text{(in)}}_-} \right]  \cr 
    &=& \delta_{\mu\nu} \left[r_h^{2a_1} e^{-\partial_{a_1}F} \frac{\sum_{\theta' = \pm} \mathcal{M}_{-\theta'} (a_h, a; a_0) \mathcal{M}_{(-\theta')+} (a,a_1; a_\infty) z_h^{\theta' a} e^{-\frac{\theta'}{2}\partial_a F}}{\sum_{\theta' = \pm} \mathcal{M}_{-\theta'} (a_h, a; a_0) \mathcal{M}_{(-\theta')-} (a,a_1; a_\infty) z_h^{\theta' a} e^{-\frac{\theta'}{2}\partial_a F}} \right] \cr 
    && \cr 
    &&
\end{eqnarray}
We get $\delta_{\mu\nu}$ in the above expression because the transverse gauge fields' equations of motion in the bulk decouple from one another. 

We notice that, the above retarded Green's function has a divergence due to the presence of $\Gamma(-1)$ in the connection matrix elements $\mathcal{M}_{++}\left(a, a_1 ; a_{\infty}\right)$ and $\mathcal{M}_{-+}\left(a, a_1 ; a_{\infty}\right)$
\begin{eqnarray}
    \left \langle J_\mu (-\omega, -\mathbf{k}) J_\nu (\omega, \mathbf{k}) \right \rangle_R = \Gamma(-1) \left \langle J_\mu (-\omega, -\mathbf{k}) J_\nu (\omega, \mathbf{k}) \right \rangle_R^{regular}
\end{eqnarray}
However the divergence can be taken care of by appropriate regularization~\cite{Bzowski:2015pba}. For example, one can consider the dimensional regularization of the gamma functions with $d= 4- \epsilon $ as
\begin{eqnarray}
    \Gamma[1-d/2] = -\frac{2}{\epsilon} + (\gamma-1) +\mathcal{O} (\epsilon)
\end{eqnarray}
where $\gamma$ is Euler's constant. 
\section{IN mode} \label{IN mode}
As we stretched out in the introduction, the identification of ingoing and outgoing modes at the asymptotic boundary of AdS spacetime is subtle because the boundary behavior of radial solution in AdS geometry turns out to be a linear combination of normalizable and non-normalizable modes. However, at the asymptotic boundary one can express it in terms of ingoing  and outgoing modes by changing  its basis of solutions \cite{Harmark:2007jy,Noda:2022zgk}. We shall briefly summarize the mathematical steps below: 
As we have seen in sections \ref{asymptotic0} and \ref{asymptotic}, the two linearly independent solutions of radial equation (2.24) in the far region can be given by  Bessel and Neumann functions, 
\begin{equation} \label{bessel_neumann}
    u^2 J_{\Delta-2} \left( \sqrt{ 1 - \frac{\mathbf{k}^2}{\omega^2} } u\right) \quad \text{and} \quad  u^2 N_{\Delta-2} \left( \sqrt{ 1 - \frac{\mathbf{k}^2}{\omega^2} } u\right)
\end{equation}
with their asymptotic behaviors given by
\begin{equation} \label{asym_bessel_neumann}
     A_J (\omega, \mathbf{k}) r^\Delta\quad \text{and} \quad A_N (\omega, \mathbf{k})  r^\Delta + B_N (\omega, \mathbf{k})  r^{4-\Delta}  
\end{equation}
Now, in order to express the solutions into the ingoing and outgoing modes, we change its basis from Bessel and Neumann functions to Hankel functions as follows:
\begin{equation} \label{hankel_fns}
    H^{(1)}_{\Delta-2}=J_{\Delta-2}+i N_{\Delta-2} \quad \text{and} \quad  H^{(2)}_{\Delta-2}=J_{\Delta-2}-i N_{\Delta-2}
\end{equation}
Note that the asymptotic form of above Hankel functions\footnote{ When $\nu$ is real, \begin{eqnarray}
    H^{(1)}_\nu (z) \sim \sqrt{\frac{2}{\pi z}} \exp \left[ i \left( z - \nu \frac{\pi}{2} - \frac{\pi}{4} \right) \right] \quad \quad \text{as} \ \  z \to \infty \\
    H^{(2)}_\nu (z) \sim \sqrt{\frac{2}{\pi z}} \exp \left[ -  i \left( z - \nu \frac{\pi}{2} - \frac{\pi}{4} \right) \right]  \quad \quad \text{as} \ \ z \to \infty 
\end{eqnarray}} can be identified with the ingoing and outgoing modes respectively. By using \eqref{asym_bessel_neumann} into \eqref{hankel_fns}, we obtain the ingoing and outgoing mode as follows:
\begin{equation} \label{ingoing_outgoing}
\begin{split}
& \text{ingoing:} \quad (A_J+iA_N)r^\Delta + i B_Nr^{4-\Delta} \equiv \Psi^{+}_{\omega,k}(r)  \\
     & \text{outgoing:} \quad (A_J-iA_N)r^\Delta - i B_Nr^{4-\Delta} \equiv \Psi^{-}_{\omega,k}(r) 
\end{split}
\end{equation}
By using \eqref{phi_decomposition}, we can write
\begin{equation}
    \Psi^{(11)}_{\omega,k}(r)=A_{11}r^\Delta \quad \text{and} \quad \Psi^{(12)}_{\omega,k}(r)=A_{12}r^{4-\Delta} 
\end{equation}
Using above expressions into \eqref{ingoing_outgoing}, we obtain:
\begin{equation}
\begin{split}
& \text{ingoing:} \quad \frac{(A_J+iA_N)}{A_{11}}\Psi^{(11)}_{\omega,k}(r) + i \frac{B_N}{A_{12}}\Psi^{(12)}_{\omega,k}(r) \equiv \Psi^{+}_{\omega,k}(r)  \\
     & \text{outgoing:} \quad \frac{(A_J-iA_N)}{A_{11}}\Psi^{(11)}_{\omega,k}(r) - i \frac{B_N}{A_{12}}\Psi^{(12)}_{\omega,k}(r)  \equiv \Psi^{-}_{\omega,k}(r) 
\end{split}
\end{equation}
Furthermore, by using \eqref{scalar_field_trans}, we can write the modes in $z$-coordinates,
\begin{equation}
\begin{split}
& \text{ingoing:} \quad \frac{(A_J+iA_N)}{A_{11}}\Psi^{(11)}_{\omega,k}(z) + i \frac{B_N}{A_{12}}\Psi^{(12)}_{\omega,k}(z) \equiv \Psi^{+}_{\omega,k}(z)  \\
     & \text{outgoing:} \quad \frac{(A_J-iA_N)}{A_{11}}\Psi^{(11)}_{\omega,k}(z) - i \frac{B_N}{A_{12}}\Psi^{(12)}_{\omega,k}(z)  \equiv \Psi^{-}_{\omega,k}(z) 
\end{split}
\end{equation}
Thus, we can explicitly define the ingoing and outgoing modes near the boundary of AdS spacetime. Moreover, it is true that the net flux at the boundary of AdS vanishes due to normalizable and non-normalizable being interpreted by real modes. However, the above construction of ingoing and outgoing modes allows to define the flux near the boundary of AdS spacetime. 

\printbibliography

\end{document}